\documentclass[10pt,twoside,reqno]{amsart}
\usepackage{mathbbol,mathtools,slashed}

\usepackage{multirow}
\usepackage[bookmarksnumbered, plainpages]{hyperref}
\usepackage[rightcaption]{sidecap}
\usepackage{caption,xparse}
\usepackage{pgfplots}

\usepackage{float}

\usepackage[english]{babel}

\usepackage[latin1]{inputenc}
\usetikzlibrary{positioning,arrows,patterns}

\usepackage{graphicx}
\usepackage{epstopdf}
\usepackage{subfig}
\usepackage{color}
\usepackage{amsthm}
\usepackage{amsmath, amsfonts, amssymb}
\usepackage[figuresright]{rotating}
\begin{document}
   \title{Real time thermal photon-photon interactions in the mixed space}
   \author{{\small M. A. A. AHMED$^{1,2,\dag}$
    H. ZAINUDDIN$^{1,3,\ddag}$, N. M. Shah$^{1,3,\star}$, and M. Ladrem$^4$}}
    \thanks{\scriptsize
 emails:\\ $^\dag$ mohammed\_h7@$_{\text{yahoo.com}}^{\text{taiz.edu.ye}}$\\$^\ddag$ hisham@upm.edu.my.\\ $^\star$risya@upm.edu.my}
 \maketitle

\address{$^{1}$ Laboratory of Computational Sciences and Mathematical Physics, Institute for Mathematical Research, Universiti Putra Malaysia, UPM Serdang 43000, Selangor, Malaysia\\
 $^{2}$Physics Department, Faculty of Science, Taiz University Al-Turba branch, Taiz, Yemen.\\
 $^{3}$ Department of Physics, Faculty of Science, Universiti Putra Malaysia, 43400 Serdang, Selangor, Malaysia.\\
 $^{4}$ Physics Department, Faculty of Science, Taibah University, Al-Madinah Al-Munawwarah, Saudi Arabia.}

\begin{abstract}
Using the effective Lagrangian for the low-energy/temperature of photon-photon interaction and the lowest-order photon self-energy is calculated in the Real Time Formalism (RTF) for an arbitrary path specified by the $\sigma$-parameter within the new basis. The causal Green's functions (without chemical potential) for the scalar field are evaluated to derive the usual thermal propagators in the mixed space. It is shown that the symmetric propagator does not depend on $\sigma-$ parameter. Furthermore, the photon self-energy is used to calculate some electromagnetic properties, such as dielectric tensor and velocity of light from photon self-energy in the mixed space that greatly simplifies calculations in RTF. Their time dependence is investigated, and a comparison between our results and those obtained by other models is discussed.
\end{abstract}
\markboth{\rightline {\sl M. A. A. Ahmed et al. }}
        {\leftline{\sl  RTF thermal photon-photon interactions in the mixed space}}

\section{\textbf{INTRODUCTION}}
The many-body problems in physics require the use of Thermal Field Theory (TFT) or the Statistical Field Theory methods. One can think, for instance, about supra-conductivity, supra-fluidity, Bose-Einstein Condensate systems, QED plasma, QCD plasma, and cosmology.  All mentioned many-body systems share some common features, meaning that the physical properties of interest depend manifestly on temperature and time. Unfortunately, the known standard framework of the QFT at zero temperature cannot be used to study the many-body phenomenon. The extension of this QFT at zero temperature within the concept of temperature and time is necessary to describe physical situations of equilibrium and non-equilibrium, which appear both in particle physics and in cosmology. Generally, only one of the fundamental interactions is responsible for the entire properties of the system. In the QED plasma, we have the electromagnetic interaction; however, the strong interaction governs all related phenomena in the QCD plasma.
The Green's functions method provides a powerful tool to evaluate the entire properties of many-body systems in both thermal equilibrium and non-equilibrium cases. It plays a fundamental role in statistical physics since they represent the bridge relating between the experimental observables and theoretical ones.
Along the extension road of the QFT techniques into the TFT ones, different formalisms developed by several authors have come into existence, that we want to recall them, succinctly: (1) Imaginary Time (IT) Formalism initiated by Matsubara \cite{Mastsubar10}. We can evaluate thermal averages if one replaces the time $t$ variable with the imaginary time variable $i\beta$ in the context of the Matsubara formalism with the help of the formal analogy between imaginary time and inverse temperature. The time integration is performed along the imaginary axis, from $0$ to $i\beta$. Therefore, Matsubara Green's functions do not contain the time variable. The time integration is replaced by a sum using the Matsubara frequencies. (2) The Closed Time Path (CTP) Formalism was initiated by Schwinger (1961) \cite{Schwinger11}, Kadanoff and Baym (1962), and Keldysh (1964) \cite{Keldysh9}. In order to consider time-dependence in the IT prescription, thus generalizing the IT method to the CTP method, one must perform analytic continuation on the complex time plane. In order to compute the real-time Green's functions, a portion along the real axis must be included in the integration path. It was developed and widely used to describe and study various related problems of non-equilibrium and equilibrium physical systems. It has also been successfully applied to the transport theory of various systems. (3) Real Time (RT) Formalism initiated by Schwinger and Keldysh. Both the CPT and RT approaches require the doubling of the degrees of freedom to consider the thermal effects. (4) Thermo Field Dynamics (TFD), initiated by Takahashi and Umezawa \cite{Takahashi12,Khanna1} in the early seventies for investigating the thermal equilibrium properties of quantum fields, has succeeded faithfully to maintain the beautiful structure of the QFT at zero temperature as is widely known. The Green's function method and the path integral technique, namely the operator formalism in TFD. The exploited idea in this sense concerns the existence of an infinite number of unitarily inequivalent representations in QFT at zero temperature. The uniqueness of the solution can be gained if one inserts the missing degrees of freedom, called by Umezawa thermal degrees of freedom. With the introduction of these degrees of freedom into QFT at $T=0$, the new QFT at finite temperature would become self-contained. The strong feature of TFD resides in the fact that quantum and thermal effects are considered in a common and in a very elegant
way, which is very convenient, especially in  the case of non-equilibrium situations with dissipation. Thus, TFD is a real time operator formalism of QFT at finite temperature, including the thermal degrees of freedom and preserving all the usual computational techniques of QFT at zero temperature, such as the Green's function technique and the invariant operator transformations.
New concepts are introduced within TFD, namely: the thermal state, a pure state rather than a mixed states, describes the thermal features of the system of interest at a given time. The thermal state is an element of the thermal space generated through the cyclic action of certain creation operators onto the thermal vacuum. Thus, one can say that the state of the system is completely determined at any time $t$ if one knows its full Hamiltonian and the thermal state at the initial time.
In the case of equilibrium situations, the thermal vacuum state is unique and depends on the heat bath temperature to which the system is coupled. However, in the case of non-equilibrium situations, it will naturally evolve in time according to the system's dynamics expressed through its Hamiltonian.
In TFT, the perturbative calculations and Feynman integrals are very similar to those in QFT at zero temperature, but the price one has to pay is a matrix structure of the propagators and self-energies, owing to the doubling of degrees of freedom. Although in principle, the full theory may be reconstructed from Green's functions, an operator formulation of TFT might be useful for practical and theoretical reasons \cite{Landsman87}. Thermal effects of some many-body systems involving photons and electrons are governed by electromagnetic interaction and the corresponding theory of Quantum Electrodynamics (QED). In QED, the mediating field is the photon field $A_{\mu}$, and its Lagrangian density possesses a $U(1)$ gauge symmetry as well as Lorentz invariance, charge conjugation symmetry, parity, and time-reversal.
Recent studies for the low temperature and high-temperature approximation for the Euler-Heisenberg effective Lagrangian, and also the low-energy effective Lagrangian below the electroweak scale, which were presented in \cite{Manjarres2017,Jenkins2018}, respectively.
Since we are interested in describing photon-photon scattering, we must include interaction terms in the Euler-Heisenberg effective interaction $\mathcal{L}_{EH}$ \cite{Heisenberg1936}.
The effective one-loop Lagrangian density of QED describes a nonlinear interaction of electromagnetic fields.
The occurrence of virtual $e^{-}e^{+}$ pairs could be identified as the physical mechanism responsible of the photon-photon interaction occurs.  Euler and Kockel succeeded to determine the leading term of the perturbation correction to the free-photon Lagrangian density $\mathcal{L}_0$ (for small field strengths and photon frequencies well below the $e^{-}e^{+}$ pair production threshold). Indeed, the effective Lagrangian describing the photon-photon interaction at low-energy \cite{Walter,Schwinger1951,DUNNE2012} is given by $\mathcal{L}_I$:
$$\mathcal{L}_{eff}=\mathcal{L}_0+\mathcal{L}_I.$$
 The one-loop Lagrangian, often called the Euler Kockel Heisenberg Lagrangian (EKH-Lagrangian) or Euler-Heisenberg Lagrangian (EH-Lagrangian), has been used to describe a variety of electromagnetic processes. It is a nonlinear Lagrangian with nonlinear corrections up to $O(m^{-4})$. To this order, there are only two such terms, consistent with all symmetries \cite{DUNNE2012,Scharnhorst2020}
\begin{equation}\label{lagra}
  \mathcal{L}_{eff}=a_0F_{\mu\nu}F^{\mu\nu}+\underbrace{\frac{\alpha^2}{m^4}\left[a_1\left(F_{\mu\nu}F^{\mu\nu}\right)^2+a_2F_{\mu\nu}F^{\nu\sigma}F_{\sigma\rho}F^{\rho\mu}\right]}\limits_{\mathcal{L}_{I}},
\end{equation}
where $F_{\mu\nu}=\partial_{\mu}A_{\nu}-\partial_{\nu}A_{\mu}$ is the electromagnetic field tensor, $a_0=-\frac{1}{2^2}$, $a_1=- \frac{1}{2^2}\frac{1}{3^2}$, and $a_2=-\frac{14}{5}a_1$, (in this case $a_1, \ a_2$ are dimensionless). The second and third parameters have to be determined from QED, where $m$ is the electron mass, and $\alpha=e^2/4\pi$ is the fine-structure constant. The resulting effective  Lagrangian does not satisfy the superposition principle and describes nonlinear electromagnetic phenomena such as photon-photon interaction. Within the framework of QED, further investigation of processes such as photon-photon scattering can be studied. Interestingly, however, the theoretical study of the propagation of light within QED only began in the 1950's (J. S. Toll)\cite{Scharnhorst2020}. The order of $m$ in $\mathcal{L}_I$ is $O(m^{-4})$. Therefore, in photon-photon interaction, a vertex can appear in any diagram at most once. The only lowest-order photon self-energy $\Pi_{\mu\nu}$, which is given by the tadpole diagram of fig(\ref{fig1}), vanishes at zero temperature. Therefore, the full photon propagator is equal to the free one \cite{Grozin2020}. Note also that the longitudinal and transverse photon self-energy within the complete one-loop approximation and the Hard Thermal Loop were studied in QED in \cite{thoma1998}, and Hard Dense Loop approximation in \cite{Stetina2018}. Indeed, at finite temperature, the EH effective theory and TFT help us to treat the photon self-energy results in a systematic way via RTF, which allows one to extend to non-equilibrium situations \cite{Thoma2000}. Based on the previous discussions, it is possible to represent the exact photon propagator in QED even with higher loop corrections. However, this requires careful consideration because the photon propagator can be gauge dependent. The propagator at finite temperature can also be computed using RTF \cite{KAPUSTA}. RTF has been widely used to study condensed matter phenomena and solve the many-body problems \cite{Khanna1,Khanna2}. The scalar propagator at finite temperature is given in many works within the momentum space. When Fourier-transforming the components of the propagator in the energy variable, we obtain the mixed space propagator for the general contour. Moreover, it is most useful to work in the mixed space where the energy variable is Fourier transformed. The propagator is a function of time and the spatial component of momentum \cite{das2018}. Such a (mixed space) representation is quite useful in many studies at finite temperature.
The mixed space representation was used in scalar field theories, which proved in a simple manner that the Feynman graphs at finite temperature are related to the corresponding zero temperature diagrams through a simple thermal operator in both ITF and RTF \cite{das2005}. The result shows that in the mixed space, the thermal propagator in RTF can be separated nicely into a zero temperature part and a finite temperature part, much like in RTF \cite{das2006a,das2006}. In this work, based upon \cite{Thoma2000}, and taking into account mixed space (momentum-time) \cite{das2018}, we carry out a covariant study of QED at finite temperature by using RTF and compare its results with those obtained by other authors in \cite{TARRACH1983,Kong1998}. We characterize the path in RTF and summarize and its photon-photon interaction, which relevant to the effective $2n$-photon vertex in a thermal photon gas where $n\geq2$ for QED in the mixed space. At finite temperature, it is noted that calculations are simplified when one works in a mixed space where energy has been Fourier transformed, making the time dependence explicit. Since time is not symmetric, working in a mixed space, where the Green's functions are dependent on time coordinate and spatial momentum \cite{DAS}, is more appropriate. Furthermore, it is now well established that temperature anisotropies in Cosmic Microwave Background (CMB) is considered the best tool to investigate the early times of our universe \cite{Staggs2018}. In such a hot early universe, the velocity of light is reduced in comparison to that of in a vacuum. It increases continuously with time as the temperature drops. Therefore, interacting photons has been taken as an application with CMB radiation.

\section{\textbf{PROPAGATOR OF BOSON GAS IN RTF WITHOUT CHEMICAL POTENTIAL}}\label{three}
Computations of Green's functions at finite temperature have undergone extensive development. However, the time and temperature-dependent formalism cannot fully accommodate the causal Green's function method, which is such a powerful technique in usual quantum field theory. On the other hand, the well-known Matsubara formalism takes advantage of Feynman diagram techniques, but it is difficult for this formalism to deal with time-dependent phenomena. Furthermore, being formulated in terms of Green's functions, the temperature Green's function methods cannot easily utilize many kinds of operator transformations. Therefore, it is very desirable to reformulate the whole structure of quantum field theory by taking into account thermal effects. Since 1963 it has become common knowledge among axiomatic field theorists that the quantum theory of free fields at finite temperature can be consistently formulated when the number of degrees of freedom is doubled \cite{Keldysh9}. In our work, we need, in most cases, to evaluate causal Green's functions (without chemical potential) in an arbitrary path for the scalar field. The propagators will be written as $2\times2$ matrices whose elements correspond to the usual thermal propagators in the mixed space $(t,\textbf{p})$ \cite{das2018,das2006}
\begin{equation}\label{propagator}
D^{(\sigma,1/2)}(t,\textbf{p};\beta)=\int_{-\infty}^{\infty}\frac{dp_0}{2\pi}\exp^{-ip_0t}D^{(\sigma,1/2)}(P;\beta),
\end{equation}
where thermal propagators in the momentum space $(p_0,p)$,
 \begin{equation}\label{pr0}
   D^{(\sigma,1/2)}(P;\beta)=\left(\begin{array}{ll}
D^{(\sigma,1/2)}_{11}(P;\beta)&D^{(\sigma,1/2)}_{12}(P;\beta)\\
D^{(\sigma,1/2)}_{21}(P;\beta)&D^{(\sigma,1/2)}_{22}(P;\beta)\end{array}\right),
\end{equation}
 where $(0\leq \sigma\leq1)$, $P$ is the incoming momentum, $T$ represents temperature and $\beta$ denotes inverse temperature.
  \begin{equation}\label{pro}
 \begin{array}{ll}
D^{(\sigma,1/2)}_{11}(P;\beta)=-i\Delta(P)-2i\pi n_B(|p_{0}|;\beta)\delta(P^2-m^2),\\
D^{(\sigma,1/2)}_{12}(P;\beta)=-2i\pi \left(\Theta(-p_0)+n_B(|p_{0}|;\beta)\right)\delta(P^2-m^2)e^{\sigma\beta p_{0}},\\
D^{(\sigma,1/2)}_{21}(P;\beta)=-2i\pi \left(\Theta(p_0)+n_B(|p_{0}|;\beta)\right)\delta(P^2-m^2)e^{-\sigma\beta p_{0}},\\
D^{(\sigma,1/2)}_{22}(P;\beta)=-i\Delta^*(P)-2i\pi n_B(|p_{0}|;\beta)\delta(P^2-m^2).
\end{array}
 \end{equation}
 The Feynman propagator $\Delta(P)$ and hence its complex conjugate $\Delta^*(P)$ is defined at zero temperature for a scalar particle as
 $$\Delta(P)=i(P^2-m^2+i\eta)^{-1}.$$
By virtue of the limit representation of the delta function
\begin{equation}\label{del1}
  \delta(P^2-m^2)=\lim_{\eta\rightarrow0}\frac{1}{\pi}\frac{\eta}{(P^2-m^2)^2+\eta^2},
\end{equation}
we can obtain,
\begin{equation}\label{del2}
  \Delta(P)+\Delta^*(P)=2i\pi \delta(P^2-m^2)=2i\pi \delta(p_0^2-E_p^2).
\end{equation}
In the following, we use $n_B(p_{0};\beta)\equiv n_B(|p_{0}|;\beta)$ and $Q^2=P^2-m^2=p_0^2-(\textbf{p}^2+m^2)$, where $Q$ is outgoing momentum. Note $D^{RTF}(Q;\beta=\infty)$ is the propagator at zero temperature $T=0$ and $D^{RTF}(Q;\beta\neq0)$ is the propagator at finite temperature $T\neq0$. That is very strong support of the theory in the sense that RTF doesn't use at all the same mathematical framework. In RTF language, the $2\times2$ matrix causal green's function is a function of the momentum space for $D^{RTF}(Q;\beta)$, where the subscripts $(1,2)$ refer to the two real branches of the time contour, these are the fields of the 1/2 basis. There are two types of components of propagators. The $``12"$ and $``21"$ components are unphysical since one of the time arguments has an imaginary component. The only physical propagators are the $``11"$ and $``22"$ components. In our work, there is one effective $4$-photon vertex. Then, one has to use eq. (\ref{pr0}), for which we ignore the indices $a$ and $b$ in the tadpole diagram through which the path $Q$ flows.
Now we will introduce a new basis representation of the RTF. It is constructed from linear combinations of the components of the RTF Green's functions \cite{Umezawa1994}. One can present (and it makes a lot of sense) the thermal propagator on the new basis. To this end, we make an orthogonal transformation \cite{Smilga1997,Ghiglieria2020}, applied to the original basis 1/2 and conducting to the new basis
\begin{equation}\label{tran}
  \mathbb{T}(\varphi)=\left(
      \begin{array}{cc}
        \cos\varphi & -\sin\varphi \\
        \sin\varphi & \cos\varphi \\
      \end{array}
    \right).
\end{equation}
Indeed, the matrix elements in this basis are built after we use some algebra to get on four components of the propagator satisfy the orthogonal transformation (\ref{tran}) brings the propagator in this form
\begin{equation}\label{tran1}
\tilde{D}^{(\sigma,1/2)}=\mathbb{T}\left(\varphi\right)D^{(\sigma,1/2)}\mathbb{T}\left(\varphi\right)^{-1},
\end{equation}
that can also be written as
\begin{equation}\label{proRA}
\tilde{D}^{(\sigma,1/2)}=\left(
      \begin{array}{cc}
        \tilde{D}^{(\sigma,1/2)}_{11} & \tilde{D}^{(\sigma,1/2)}_{12} \\
        \tilde{D}^{(\sigma,1/2)}_{21} & \tilde{D}^{(\sigma,1/2)}_{22} \\
      \end{array}
    \right),
\end{equation}
then
\begin{equation}\label{rotation}
  \begin{array}{l}
\tilde{D}^{(\sigma,1/2)}_{11}=D^{(\sigma,1/2)}_{11}+\sin^2\varphi (D^{(\sigma,1/2)}_{22}-D^{(\sigma,1/2)}_{11})-\frac{1}{2}\sin2\varphi(D^{(\sigma,1/2)}_{12}+D^{(\sigma,1/2)}_{21}),\\
\tilde{D}^{(\sigma,1/2)}_{12}=D^{(\sigma,1/2)}_{12}-\sin^2\varphi (D^{(\sigma,1/2)}_{12}+D^{(\sigma,1/2)}_{21})+\frac{1}{2}\sin2\varphi(D^{(\sigma,1/2)}_{11}-D^{(\sigma,1/2)}_{22}),\\
\tilde{D}^{(\sigma,1/2)}_{21}=D^{(\sigma,1/2)}_{21}-\sin^2\varphi (D^{(\sigma,1/2)}_{12}+D^{(\sigma,1/2)}_{21})+\frac{1}{2}\sin2\varphi(D^{(\sigma,1/2)}_{11}-D^{(\sigma,1/2)}_{22}),\\
\tilde{D}^{(\sigma,1/2)}_{22}=D^{(\sigma,1/2)}_{22}+\sin^2\varphi (D^{(\sigma,1/2)}_{11}-D^{(\sigma,1/2)}_{22})+\frac{1}{2}\sin2\varphi(D^{(\sigma,1/2)}_{12}+D^{(\sigma,1/2)}_{21}),
\end{array}
\end{equation}
the components ``11'' and ``22'' are called the symmetric propagator, and the components ``12'' and ``21'' are called the mixed propagators. The new components of the propagator in momentum space are,
\begin{equation}\label{rraa}
  \begin{array}{l}
\tilde{D}^{(\sigma,1/2)}_{11}(Q;\beta)=\frac{1}{i}\left[\Delta(Q)+2\pi\delta(Q^2)n_B(Q;\beta)\right]-\mathcal{P}\left(\frac{1}{Q^2}\right)\sin^2\varphi\\
\ \ \ \ \ \ \ \ \ \ \ \ \ \ \ \ \ \ \ \ -\frac{\pi\delta(Q^2)}{i}\left(\Theta(-q_0)e^{\sigma\beta q_0}+\Theta(q_0)e^{-\sigma\beta q_0}+2n_B(Q;\beta)\cosh\sigma\beta q_0\right)\sin2\varphi,\\
\tilde{D}^{(\sigma,1/2)}_{12}(Q;\beta)=\frac{2\pi\delta(Q^2)}{i}\left[\left(\Theta(-q_0)+n_B(|q_{0}|;\beta)\right)e^{\sigma\beta q_{0}}\right]+\frac{1}{2}\mathcal{P}\left(\frac{1}{Q^2}\right)\sin2\varphi \\
\ \ \ \ \ \ \ \ \ \ \ \ \ \ \ \ \ \ \ \ -\frac{\pi\delta(Q^2)}{i}\left(\Theta(-q_0)e^{\sigma\beta q_0}+\Theta(q_0)e^{-\sigma\beta q_0}-2n_B(Q;\beta)\sinh\sigma\beta q_0\right)\sin^2\varphi,\\
\tilde{D}^{(\sigma,1/2)}_{21}(Q;\beta)=\frac{2\pi\delta(Q^2)}{i}\left[\left(\Theta(q_0)+n_B(|q_{0}|;\beta)\right)e^{-\sigma\beta q_{0}}\right]+\frac{1}{2}\mathcal{P}\left(\frac{1}{Q^2}\right)\sin2\varphi \\
\ \ \ \ \ \ \ \ \ \ \ \ \ \ \ \ \ \ \ \ -\frac{\pi\delta(Q^2)}{i}\left(\Theta(-q_0)e^{\sigma\beta q_0}+\Theta(q_0)e^{-\sigma\beta q_0}-2n_B(Q;\beta)\sinh\sigma\beta q_0\right)\sin^2\varphi,\\
\tilde{D}^{(\sigma,1/2)}_{22}(Q;\beta)=\frac{1}{i}\left[\Delta^*(Q)+2\pi\delta(Q^2)n_B(Q;\beta)\right]+\mathcal{P}\left(\frac{1}{Q^2}\right)\sin^2\varphi\\
\ \ \ \ \ \ \ \ \ \ \ \ \ \ \ \ \ \ \ \ +\frac{\pi\delta(Q^2)}{i}\left(\Theta(-q_0)e^{\sigma\beta q_0}+\Theta(q_0)e^{-\sigma\beta q_0}+2n_B(Q;\beta)\cosh\sigma\beta q_0\right)\sin2\varphi,
\end{array}
\end{equation}
where $\mathcal{P}\frac{1}{Q^2}$ is the Cauchy principal value. Here, we make substantial use of the results contained in Ref. \cite{das2018}, which give the elements propagator (\ref{pro}) in the mixed space. Hence, we recompute the propagators (\ref{rraa}) in the mixed space as
 \begin{equation}\label{rraamix}
  \begin{array}{l}
\tilde{D}^{(\sigma,1/2)}_{11}(t,q;\beta)=\frac{1}{2iq}\left[\Theta(t)e^{-iqt}+\Theta(-t)e^{iqt}+2n_B(q;\beta)\cos qt\right]+\frac{\sin qt}{q} \left(\Theta(-t)-\Theta(t)\right)\sin^2\varphi\\
\ \ \ \ \ \ \ \ \ \ \ \ \ \ \ \ \ \ \ \ -\frac{\cos qt}{2iq}\left(e^{-\sigma\beta q}+2n_B(q;\beta)\cosh\sigma\beta q\right)\sin2\varphi,\\
\tilde{D}^{(\sigma,1/2)}_{12}(t,q;\beta)=\frac{1}{2iq}\left(e^{iq(t+i\sigma\beta)}+2n_B(q;\beta)\cos q(t+i\sigma\beta)\right)+\frac{\sin qt}{q} \left(\Theta(-t)-\Theta(t)\right)\sin2\varphi\\
\ \ \ \ \ \ \ \ \ \ \ \ \ \ \ \ \ \ \ \ -\frac{\cos qt}{iq}\left(e^{-\sigma\beta q}+2n_B(q;\beta)\cosh\sigma\beta q\right)\sin^2\varphi,\\
\tilde{D}^{(\sigma,1/2)}_{21}(t,q;\beta)=\frac{1}{2iq}\left(e^{-iq(t-i\sigma\beta)}+2n_B(q;\beta)\cos q(t-i\sigma\beta)\right)+\frac{\sin qt}{q} \left(\Theta(-t)-\Theta(t)\right)\sin2\varphi\\
\ \ \ \ \ \ \ \ \ \ \ \ \ \ \ \ \ \ \ \ -\frac{\cos qt}{iq}\left(e^{-\sigma\beta q}+2n_B(q;\beta)\cosh\sigma\beta q\right)\sin^2\varphi,\\
\tilde{D}^{(\sigma,1/2)}_{22}(t,q;\beta)=\frac{1}{2iq}\left[\left(\Theta(t)e^{iqt}+\Theta(-t)e^{-iqt}\right)+2n_B(q;\beta)\cos qt\right]+\frac{\sin qt}{q} \left(\Theta(t)-\Theta(-t)\right)\sin^2\varphi\\
\ \ \ \ \ \ \ \ \ \ \ \ \ \ \ \ \ \ \ \ +\frac{\cos qt}{2iq}\left(e^{-\sigma\beta q_0}+2n_B(q;\beta)\cosh\sigma\beta q\right)\sin2\varphi.
\end{array}
\end{equation}
Now, one can also define the symmetric propagator in the mixed space as:
\begin{equation}\label{Sp}
\boxed{\tilde{D}^{(\sigma,1/2)}_S (t,q;\beta)=Tr(\tilde{D}^{(\sigma,1/2)}(t,q;\beta))=\frac{\cos qt}{iq}\left[1+2 n_B(q|;\beta)\right]}.
\end{equation}
There are several things to note from the structures of the terms of the propagator in (\ref{rraamix}). Some of these terms are independent of temperature, present at $T=0$, while other terms are temperature-dependent $T\neq0$. However, all components of the propagator are dependent on the arbitrary $\sigma$-parameter characterizing the path chosen in the new basis. This dependence can be cancelled in the case of the particular value of $\varphi$-transformation and also for the symmetric propagator, as we have just noted above.

We would like to say when we choose the angle $\varphi=0$, we get the old basis, moreover the old basis with a negative sign is obtained when $\varphi=90$. Therefore, we choose $\varphi=45$ leading to a symmetric propagator, and to the appearance of the $\sigma$ parameter in each component of propagator in the new basis as we see in table 1. The two popular choices for the parameter $\sigma$ appearing in the Feynman rules in eq.(4) for the propagator are $\sigma=0$ and $\sigma=\frac{1}{2}$. The first case is known as closed-time path (CTP) formalism by Schwinger-Keldysh, and the second case is known as thermofield dynamics (TFD) by Takahashi-Umezawa \cite{Gozzi2011,DAS2016}.

\section{\textbf{THERMAL INTERACTION OF PHOTONS GAS IN RTF}}\label{four}
\begin{figure}[!h]
  \centering
 \includegraphics[width=380pt]{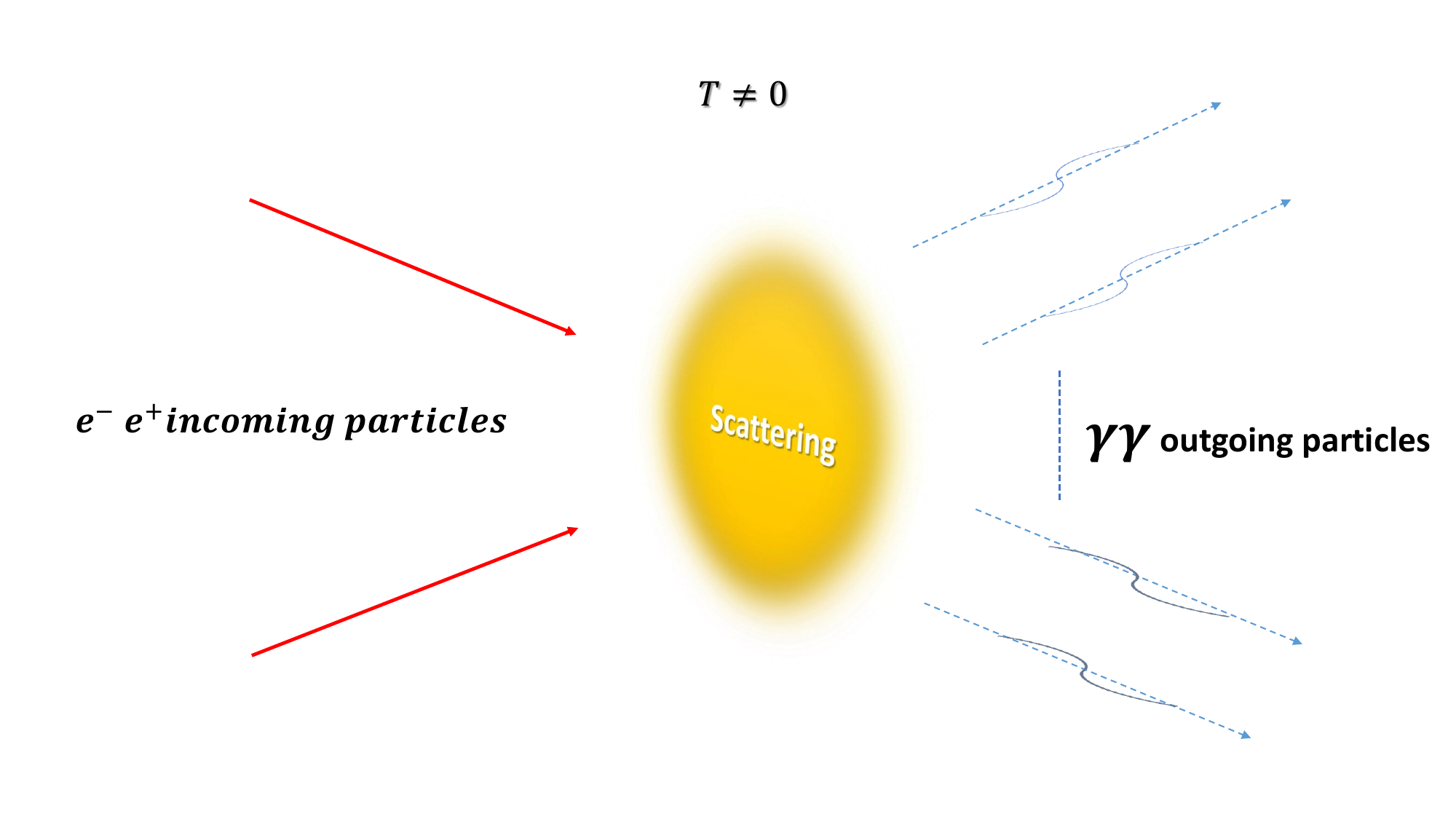}
 \captionsetup{singlelinecheck=off}
  \caption{$e^-e^+ \rightarrow$ photon-photon scattering process}\label{fig0}
\end{figure}

\begin{figure}[!h]
  \centering
 \includegraphics[width=250pt]{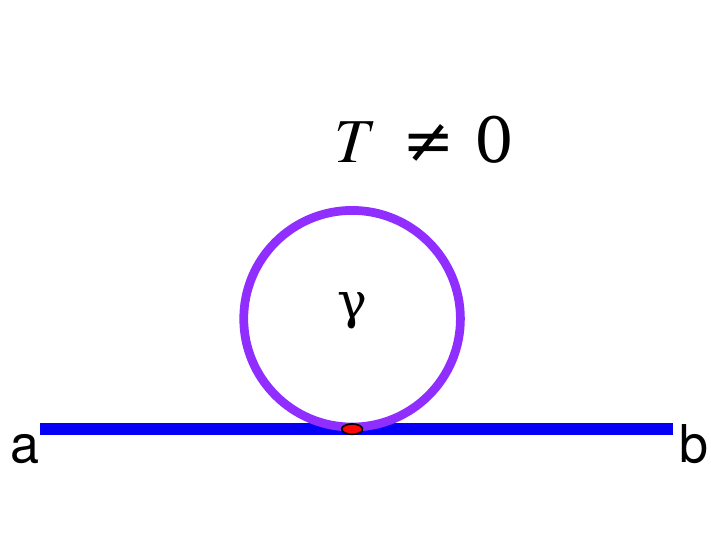}
  \caption{Tadpole diagram: The red blob denotes the effective photon vertex  (retarded vertice)
}\label{fig1}
\end{figure}

The self-energy of photons in the tadpole diagram is an important example of the application of QED perturbation theory, and it does not violate the gauge symmetry of QED at finite temperature. In fig.(\ref{fig0}), we consider a thermal system of relativistic electrons $e^-$, positrons $e^+$, and photons with a net charge zero, i.e. with vanishing chemical potential. In the RTF, there is a one-parameter family of paths (for any value of the arbitrary parameter $\sigma$). Therefore we use RTF in this work. Of interest are the dispersion relations of photons in a thermal photon gas and its dielectric function, related to the index of refraction and the velocity of light in thermal mixed space \cite{das2018}.
 In this work, we consider the photon self-energy tensor in the effective theory for photon-photon interaction of low energy photons. The photon self-energy tensor $\tilde{\Pi}_{\mu\nu}$ is a ($4\times4$) matrix, describes the propagation of photons in 4-dimensional space-time and carries the information of polarization. Transversality of the photon is associated when $m_{\gamma}=0$ as a gauge condition and is related to the absence of interaction of photons with the medium, leading to the absence of its longitudinal component.
In fig (\ref{fig1}) we find for the dressed one-loop contribution to the retarded photon self-energy in RTF which given by
\begin{equation}\label{selfenergy}
  \tilde{\Pi}_{\mu\nu}(P;\beta)=-\frac{1}{2}\int\frac{d^dQ}{(4\pi)^d}\mathcal{D}^{\beta\rho}(Q;\beta)\mathcal{T}_{\rho\mu\beta\nu}(Q,P,Q,P),
\end{equation}
where $\mathcal{T}_{\rho\mu\beta\nu}(Q,P,Q,P)$ is the effective photon vertex. In the tadpole diagram in fig.(\ref{fig1}), there are four ways to attach the first leg of the photon vertex to $a$, and three ways to attach the second leg to $b$. Hence the weight $\frac{1}{4!}4\times3=\frac{1}{2}$ is the symmetry factor associated with the tadpole diagram. In a covariant gauge \cite{Carignano2018}, the photon propagator is given by,
\begin{equation}\label{prowhole}
  \mathcal{D}^{\beta\rho}(Q;\beta)=\frac{1}{2} (i^2)\left[ g^{\beta\rho}-\kappa \hat{Q}^{\rho}\hat{Q}^{\beta}\right]\tilde{D}^{(\sigma,1/2)}(Q;\beta),
\end{equation}
and the gauge parameter $\kappa$ in (\ref{prowhole}) is especially convenient for perturbative calculations as it can minimize the number of terms in the photon propagator \cite{Carignano2018}, and was also used in \cite{Thoma2000}.
Let $Q^{\rho}$ be the momentum tensor of the photon. Current conservation requires that $\mathcal{T}_{\rho\mu\beta\nu}$ obeys
\begin{equation}\label{wt}
  Q^{\rho}Q^{\beta}\mathcal{T}_{\rho\mu\beta\nu}(Q,P,Q,P)=0,
\end{equation}
where $Q$ is the external momentum in the tadpole diagram, and $P$ is the loop momentum. Gauge invariance requires that the parameter $\kappa$ does not appear and that this constraint holds at $T>0$, $\kappa\neq0$, as well as in the vacuum \cite{KAPUSTA}.
The effective photon vertex of $\mathcal{L}_{I}$ in (\ref{lagra}) is $\mathcal{T}^{\rho\beta\mu\nu}(Q,P,Q,P)$ for the lowest order of photon-photon interaction. With some factors of 2 and the sign corrected \cite{Dicus1998}, and adopting the expression for the four-photon vertex given in \cite{Halter1993} (see \ref{a}), we obtain the transverse component $(ij)$,
\begin{equation}\label{tenxx}
\mathcal{P}^{ij}\mathcal{T}^\rho_{i\rho j}(Q,P,Q,P)\Big|_{Q^2=0}=i\left(4\frac{\alpha}{m^2}\right)^2(4a_1+3a_2)\left(p_0^2+ p^2\right)q^2(\cos^2\theta +1),
\end{equation}
where $\mathcal{P}^{ij}=\frac{1}{d-2}(\delta^{ij}-\frac{p^i p^j}{p^2})\ (i,j=1,..,3)$ is the appropriate projection operator. The longitudinal component is given by,
\begin{equation}\label{ten00}
\mathcal{T}^\rho_{0\rho 0}(Q,P,Q,P)\Big|_{Q^2=0}=-i\left(4\frac{p\alpha}{m^2}\right)^2(4a_1+3a_2)q^2(\cos^2 \theta -1)\ .
\end{equation}
\section{\textbf{RESULTS AND DISCUSSION}}\label{five}
In the previous section, we introduced the full propagator of photon gas and the lowest order photon self-energy. As well as there are only two independent components of $\tilde{\Pi}_{\mu\nu}$, that depend on $p_0$ and $p$ for which we choose $(ij)$ or/and $(00)$, which is given by the tadpole in Fig(\ref{fig1}). Therefore the lowest order photon self-energy vanishes at zero temperature $T=0$ after dimensional regularization \cite{Halter1993}. However, at finite temperature, $T\neq0$ tadpole diagrams lead to a finite result \cite{KAPUSTA}. We have assumed dimensional regularization to get rid of the vacuum contributions, and we use the Minkowski metric, $P^2=P_{\mu}P^{\mu}=g_{\mu\nu}P^{\nu}P^{\mu}=p_0^2-p^2$. Four momenta are denoted by $P\equiv(p_0,p)$, for the photon self-energy. From (\ref{selfenergy}) we want to compute $\Pi_L(P)^{QED}$ and $\Pi_T(P)^{QED}$ through the matrix,
\begin{equation}\label{lotr}
 \tilde{ \Pi}_{\mu\nu}(P)=\left(
                    \begin{array}{cc}
                      \Pi_{00} & \Pi_{0j} \\
                      \Pi_{i0} & \Pi_{ij} \\
                    \end{array}
                  \right),
\end{equation}
for which $\Pi_{L}^{QED}=\Pi_{00}^{QED}$ and $\Pi_{T}^{QED}=\mathcal{P}^{ij}\Pi_{ij}^{QED}$, are the longitudinal and transverse projections, respectively. We construct the photon self-energy tensor from the photon propagator in (\ref{prowhole}) and effective photon vertex $\mathcal{T}_{\rho\mu\beta\nu}(Q,P,Q,P)$, where $\mathcal{T}_{\rho\mu\beta\nu}(Q,P,Q,P)$ is gotten by tracing over one pair of indices in the four vertex for the components of the polarization tensor in (\ref{tenxx}) and (\ref{ten00}). Then we insert (\ref{prowhole}) and (\ref{tenxx}) into (\ref{selfenergy}) to compute the transverse component, and by the same technique to compute the longitudinal component by inserting (\ref{prowhole}) and (\ref{ten00}) into (\ref{selfenergy}).
 Writing out the integral $\tilde{\Pi}^{(\sigma)}_{L}(P,\beta)$
\begin{equation}\label{long}
  \tilde{\Pi}^{(\sigma)}_{L}(P;\beta)=-\frac{1}{2}\int\frac{d^dQ}{(2\pi)^d}\mathcal{D}^{00}(Q;\beta) \mathcal{T}^{\rho}_{0\rho0}(Q,P,Q,P),
\end{equation}
and $\tilde{\Pi}^{(\sigma)}_T(P;\beta)$ is the transverse projector, is defined as
\begin{equation}\label{trens}
  \tilde{\Pi}^{(\sigma)}_{T}(P;\beta)=-\frac{1}{2}\int\frac{d^dQ}{(2\pi)^d}\mathcal{D}^{ij}(Q;\beta) \mathcal{T}^{\rho}_{i\rho j}(Q,P,Q,P),
\end{equation}
where $d$ is dimensionality of space. In our case, $d=4$, and when the system in equilibrium state, then $n_B(|q_0|,q;\beta)=\frac{1}{e^{\beta |q_0|}-1}$. However for out of equilibrium photon distribution, $n_B$ is dependent on the momentum and the space-time coordinates. At this point, we can now compute $\tilde{\Pi}_{L,T}(P;\beta)$,
\begin{equation}\label{long1}
 \tilde{\Pi}^{(\sigma)}_{L}(P;\beta)=-\frac{1}{2}\underbrace{\int\frac{d^4Q}{(2\pi)^4}\mathcal{D}^{00}(Q) \mathcal{T}^{\rho}_{0\rho0}(Q,P,Q,P)}\limits_{I_L(P;0)} -\frac{1}{2}\underbrace{\int\frac{d^4Q}{(2\pi)^4}\mathcal{D}^{00}(Q;\beta)\mathcal{T}^{\rho}_{0\rho0}(Q,P,Q,P)}\limits_{I_L(P;\beta)},
\end{equation}
and
\begin{equation}\label{trens1}
  \tilde{\Pi}^{(\sigma)}_{T}(P;\beta)=-\frac{1}{2}\underbrace{\int\frac{d^4Q}{(2\pi)^4}\mathcal{D}^{ij}(Q)\mathcal{T}^{\rho}_{i\rho j}(Q,P,Q,P)}\limits_{I_T(P;0)} -\frac{1}{2}\underbrace{\int\frac{d^4Q}{(2\pi)^4}\mathcal{D}^{ij}(Q;\beta)\mathcal{T}^{\rho}_{i\rho j}(Q,P,Q,P)}\limits_{I_T(P;\beta)}\ .
\end{equation}
 After inserting (\ref{tenxx}) and (\ref{ten00} ) into (\ref{long1}) and (\ref{trens1}), we get two kinds of integrations; one of them at zero-temperature when the thermal contributions vanishing,
\begin{equation}\label{int1}
I_L(P;0)=\left(\frac{44\alpha^2}{45 m^4}\right)p^2\int\frac{d^4Q}{(2\pi)^4}[\Delta(Q)+\pi\delta(Q^2)][q^2(\cos^2 \theta-1)],
\end{equation}
and
\begin{equation}\label{int2}
I_T(P;0)= -\left(\frac{22\alpha^2}{45 m^4}\right)(p_0^2+p^2)\int\frac{d^4Q}{(2\pi)^4}[\Delta(Q)+\pi\delta(Q^2)][q^2(\cos^2 \theta+1)]\ .
\end{equation}
To study the thermal contribution at finite temperatures, the integrals are computed in a similar way, yielding,
\begin{equation}\label{intthermal1}
I^{(\sigma)}_L(P;\beta)=i\left(\frac{44\alpha^2}{45 m^4}\right)p^2\int\frac{d^4Q}{(2\pi)^3}\left[\tilde{D}^{(\sigma,1/2)}(Q;\beta)\right]\left[q^2(\cos^2 \theta-1)\right],
\end{equation}
and
\begin{equation}\label{intthermal2}
I^{(\sigma)}_T(P;\beta)= -i\left(\frac{22\alpha^2}{45 m^4}\right)(p_0^2+p^2)\int\frac{d^4Q}{(2\pi)^3}\left[\tilde{D}^{(\sigma,1/2)}(Q;\beta)\right]\left[q^2(\cos^2 \theta+1)\right]\ .
\end{equation}
In the mixed space, where the propagators are defined as functions of the time coordinate and the spatial momentum \cite{das2018,das2006}. It is to be noted here that even though the complete Green's functions will not be time translation invariant, and consequently, the initial physical Green's functions in (\ref{rraamix}) depend only on time coordinate. At this point, one can use the propagator given in (\ref{rraamix}) in a straightforward manner. Therefore we need to choose the rotation angle of the new basis, let it be $45^\circ$. After integrating over $q_0$ in the mixed space by means of the $\delta$-functions in (\ref{intthermal1} and \ref{intthermal2}) and arrive at,
\begin{equation}\label{seinmix}
\tilde{\Pi}^{(\sigma,1/2)}_{T,L}(t,\textbf{p};\beta)= \chi_{T,L} \int_{0}^{\infty} dq\ q^4\ i \tilde{D}^{(\sigma,1/2)} (t,q;\beta)\ .
\end{equation}
Using the polygamma function definition and its series representation in our case as given by the above formula,
$$\int_{0}^{\infty}dq\frac{q^3}{e^{\beta q}-1}\cos\left(qt\right)=\frac{1}{2}\frac{\psi^3\left(1+i\frac{t}{\beta}\right)+\psi^3\left(1-i\frac{t}{\beta}\right)}{\beta^4}$$
and with ignoring the terms without temperature of $\tilde{\Pi}^{(\sigma,1/2)}_{T,L}(t,q;\beta)$ we get the results summarized in the following table,
\begin{table}[H]

\begin{tabular}{||l|c|c|c||}
  \hline
  / &$\sigma$& $\sigma=0$ &$\sigma=\frac{1}{2}$ \\
  \hline
  \hline
  $\tilde{\Pi}_{T,L}^{(\sigma,11)}(t,p;\beta)$ & $\frac{1}{4}\left(2\Psi(\tau,\bar{\tau})-\Psi^{(+)}(\tau\pm\sigma,\bar{\tau}\pm\sigma)\right)$ & $0$ &$\frac{1}{2}\left[\Psi(\tau,\bar{\tau})-\Psi\left(\tau-\frac{1}{2},\bar{\tau}-\frac{1}{2}\right)\right] $ \\

 \hline
  $\tilde{\Pi}^{(\sigma,12)}(t,p;\beta)$ & $-\frac{1}{4}\Psi^{(-)}(\tau\pm\sigma,\bar{\tau}\pm\sigma)$ & $0$ & $0$\\

  \hline
  $\tilde{\Pi}_{T,L}^{(\sigma,21)}(t,p;\beta)$ & $\frac{1}{4}\Psi^{(-)}(\tau\pm\sigma,\bar{\tau}\pm\sigma)$&$0$ & $0$ \\

  \hline
  $\tilde{\Pi}_{T,L}^{(\sigma,22)}(t,p;\beta)$ & $\frac{1}{4}\left(2\Psi(\tau,\bar{\tau})+\Psi^{(+)}(\tau\pm\sigma,\bar{\tau}\pm\sigma)\right)$ & $\Psi(\tau,\bar{\tau})$ & $\frac{1}{2}\left[\Psi(\tau,\bar{\tau})+\Psi\left(\tau-\frac{1}{2},\bar{\tau}-\frac{1}{2}\right)\right] $\\
\hline
\hline
  $\tilde{\Pi}_{T,L}^{(\sigma,S)}(t,p;\beta)$ & $\Psi(\tau,\bar{\tau})$ & $\Psi(\tau,\bar{\tau})$ & $\Psi(\tau,\bar{\tau})$\\

\hline
  \hline
  \end{tabular}
  \caption*{Table 1: The thermal photon self-energy for an arbitrary path in the mixed space for each components in the new basis, where
$\Psi(\tau,\bar{\tau})=\frac{\chi_{T,L}}{\beta^4}\left[\psi^3(\bar{\tau})+\psi^3(\tau)\right]$.}\label{tab:b}
\end{table}

\noindent where $\psi^3(\tau)$ is the PolyGamma function and $\tau=\left(1+i\frac{t}{\beta}\right)$ is a complex function on $(t;\beta)$, $\bar{\tau}$ is the conjugate function of $\tau$ \cite{Matone2006,Dittrich2019}. As we mentioned, the value of transformation angle to the new basis is $\varphi=45$, this choice allowed to appear the parameter $\sigma$ in each component of the propagator in the new basis.
Our strategy depends on two tasks; namely, this is the first time the RTF is used to introduce the scalar propagator in the mixed space (\ref{rraamix}). Furthermore, this strategy allows us to construct the photon self-energy components in the mixed space from (\ref{rraamix}) within the general path. Note that the diagonal components of photon self-energy in the new basis are not the same in both approaches within RTF, whereas the off-diagonal components of photon self-energy are vanishing in table \ref{tab:b}. This means with the new basis, it allows one to reduce the components.
 However, we have to choose the value of the parameter $\sigma$, and we stick to the popular choices. The two popular choices for $\sigma$-parameter appearing in the Feynman rules for the propagator are $(\sigma=0,\frac{1}{2})$ \cite{DAS,Gozzi2011}. It is also the original one adopted by Schwinger-Keldysh and Takahashi-Umezawa in their approaches within RTF. It is easy to find that the symmetric propagator from (\ref{rraa}) is equal to $D_{22}^{(\sigma=0,1/2)}$, hence, the results will be the same as Thoma result's in \cite{Thoma2000}. However, that allows us to study $\tilde{\Pi}^{(\sigma,S)}_{T,L}(t,p;\beta)$ which need to enter only the symmetric propagator in the mixed space,
\begin{equation}\label{mixlall}
  \tilde{\Pi}^{(\sigma,S)}_{T,L}(t,\textbf{p};\beta)=\chi_{T,L}\left(\frac{\psi^3(\tau)+\psi^3(\bar{\tau})}{\beta^4}\right),
\end{equation}
where the functions $\chi_{T,L}(p,p_0)$ refer to the transverse projector $\chi_{T}$ and longitudinal projector $\chi_{L}$ as:
\begin{equation*}
  \chi_{T,L}(p,p_0 )=\begin{cases}
                                        \frac{44}{135}\left(i\frac{\alpha p}{\pi m^2}\right)^2,  \\
                                        \\
                                        \frac{44}{135}\left(i\frac{\alpha p}{\pi m^2}\right)^2\frac{\left(p_0^2+ p^2\right)}{p^2} .\\
                                      \end{cases}
\end{equation*}
Observe that our results carry mixed dependence, i.e. on the thermal effect of the temperature and the dynamic effect of time. The speed of light at finite temperature is known to be below the ordinary vacuum value due to the interactions of the photons with virtual electron-positron pairs as shown in \cite{TARRACH1983}. The novelty that emerges is that the presence of a cold heat bath is at the origin of the variation in the speed of light, in particular its increase around the minimum point $t_{min}$ which disappears with the increase in temperature. The negative value of the function $\omega(t;T)$ is responsible for the drastic increase in the speed of light which is probably related to the low energy / temperature approximation of the model used. The interaction effects are considered here for $\tilde{\Pi}^{(\sigma,1/2)}_{T,L}(t,\textbf{p};\beta)$ using RTF which leads to the results given above. Furthermore, we ignore the non-thermal part, given that we are studying thermal interactions of photon self-energy in RTF, the results are similar in both approaches CTP and TFD for $\tilde{\Pi}^{\left(\left(\sigma=0,\frac{1}{2}\right),S\right)}_{T,L}(t,\textbf{p};\beta)$, while the thermal part in off-diagonal components of the photon self-energy is vanishing (see table \ref{tab:b}) \cite{TARRACH1983,Thoma2000,Kong1998}. In the context of TFD , we have not found any work containing the speed of light result. However, our strategy introduced the general path in RTF, which allows us to select any path within RTF, such that the path of TFD is one of them. Now let us define a two-variable function $\Omega(t;\beta)$:
\begin{equation}\label{plot}
  \Omega(t;\beta)=-\frac{44}{135}\left(\frac{\alpha}{\pi}\right)^2\left(\frac{T}{m}\right)^4\left[\psi^3(\tau)+\psi^3(\bar{\tau})\right]=-c_{TQED}\ \omega(t;\beta)\ ,
\end{equation}
where $c_{TQED}=\frac{44}{135}\left(\frac{\alpha}{\pi\ m^2}\right)^2=\frac{1.42293\times10^{-45}}{(K)^4}$ is a constant. When $t=0$, we have,
  $$\Omega(t=0;\beta)=-c_{TQED}\ \omega(t=0;\beta)=-\frac{2}{15}\pi^4c_{TQED}T^4 \ .$$
Now we rewrite $\Pi_{L,T}(t,\textbf{p};\beta)$ using $\Omega(t;\beta)$:
\begin{equation}\label{mixtherml3}
  \begin{array}{l}
   \tilde{\Pi}_{L}(t,\textbf{p};\beta) =p^2\Omega(t;\beta) \ \ \ \ \ \textrm{and} \ \ \ \ \ \tilde{\Pi}_{T}(t,\textbf{p};\beta)=(p_0^2+p^2)\Omega(t;\beta).
  \end{array}
\end{equation}
Note that the first factor in the final results of $\tilde{\Pi}_{L,T}(t,\textbf{p};\beta)$ does not depend on temperature and the second factor depends simultaneously on temperature $T=\frac{1}{\beta}$ and time $t$. That is, indeed, exactly what is obtained by solving the differential equations for the propagator in the present case.
\begin{figure}[!h]
  \centering
 \includegraphics[width=350pt]{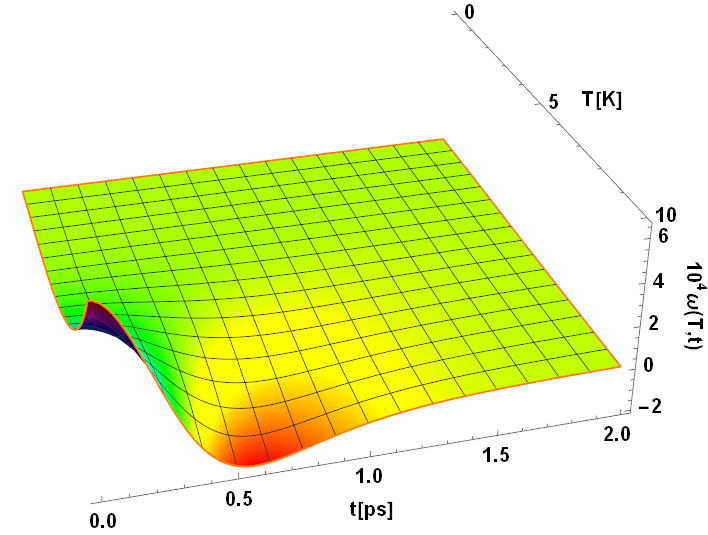}
 \captionsetup{singlelinecheck=off}
  \caption{The 3-Dim plot of $\omega(t;T)$ versus Temperature and time.}\label{figappd}
\end{figure}

\begin{figure}[!h]
  \centering
 \includegraphics[width=350pt]{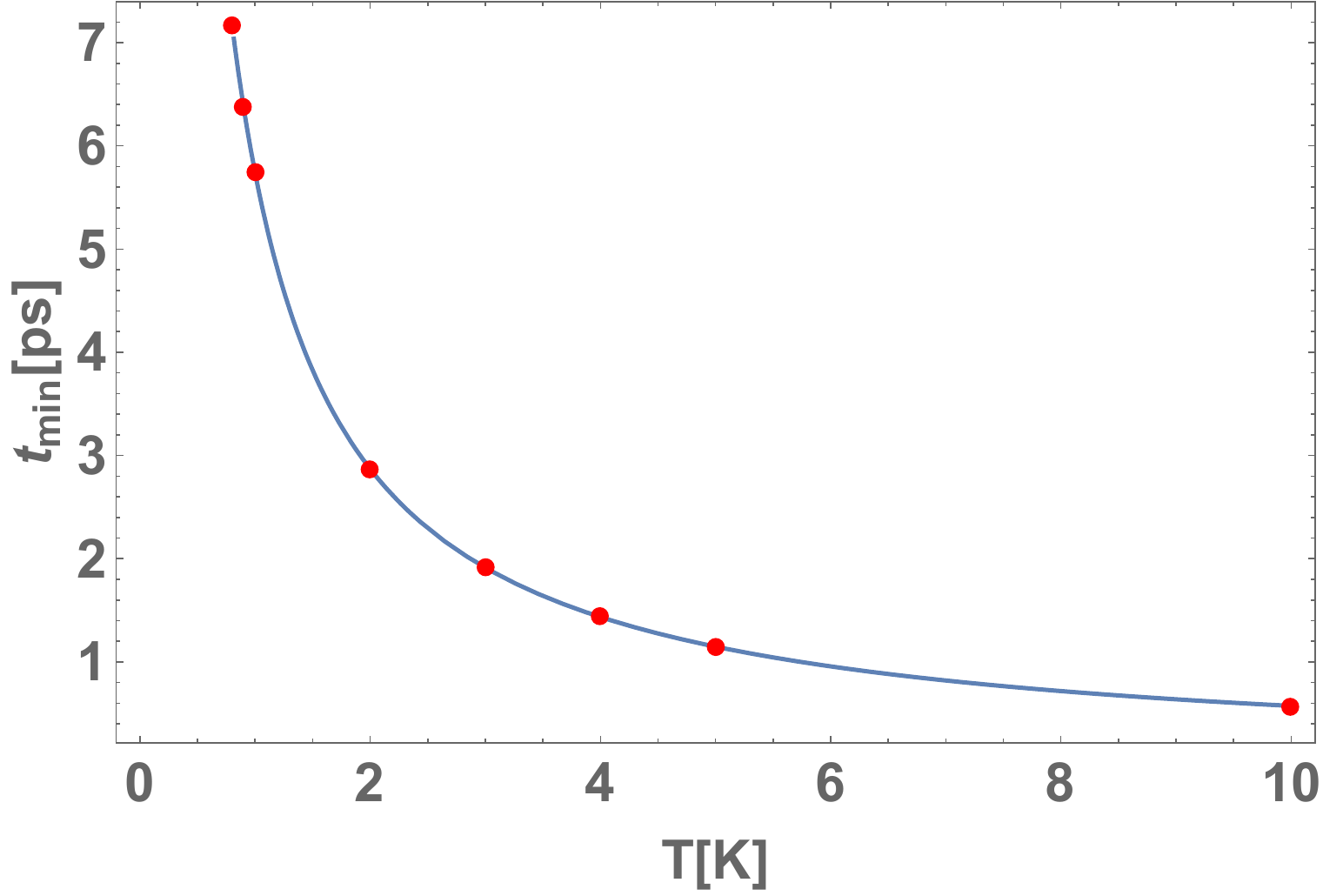}
 \captionsetup{singlelinecheck=off}
  \caption{The plot of $t_{min}(T)$ versus temperature with the fit $t_{min}(T)=5.73973 T^{-1}$ .}\label{tminvsT1}
\end{figure}

The 3-Dim plot displayed on the figure (\ref{figappd}) shows the time-Temperature behaviour of $\omega(T,t)$ at low values of temperature and time. One can notice that when $0<T<10K$ and $0<t<1ps$ the obtained values of $\omega(T,t)$ lie between $-2\times10^{-4}$ and $6\times10^{-4}$. One can clearly see the asymptotic vanishing values expressing the standard values attained by the different electromagnetic properties. However, a little oscillatory variation is noticed at the beginning of time, leading to a corresponding fluctuation of these quantities with numerical values below and above the standard known values. A minimum value arises during the time fluctuation, with a position shifted towards low values as the temperature is increased. From the 3-Dim plot, one can extract the following values: $t_{min}=1.1479\ ps$ at $T=5K$ and $t_{min}=0.5739\ ps$ at $T=10\ K$. In other words, more is high the temperature more the duration of the time fluctuation of $\omega(T,t)$ is small before reaching the asymptotic standard limit.
An interesting 2-Dim plot given by the figure (\ref{tminvsT1}), representing $t_{min}$ versus the temperature $T$, shows the decreasing behavior of $t_{min}$ as function of $T$, which can parameterized by the following relationship $ t_{min}\propto T^{-1}$.

The dielectric tensor $\varepsilon_{L,T}(t,\textbf{p};\beta)$ is closely related to photon self-energy $\tilde{\Pi}_{L,T}(t,\textbf{p};\beta)$ by \cite{Elze1989,Weldon1982}:
\begin{equation}\label{dl1}
  \varepsilon_L(t,\textbf{p};\beta)=1-\frac{\tilde{\Pi}_L(t,\textbf{p};\beta)}{p^2}=1-\Omega(t;\beta),
\end{equation}
and
\begin{equation}\label{dti}
  \varepsilon_T(t,\textbf{p};\beta)=1-\frac{\tilde{\Pi}_T(t,\textbf{p};\beta)}{p_0^2}=1-\frac{p^2+p_0^2}{p_0^2}\Omega(t;\beta).
\end{equation}
Two electromagnetic properties of the system, namely the electric permittivity and the magnetic permeability, which are independent on $p$ and $p_0$, are given by
\begin{equation}\label{ele}
  \epsilon(t;\beta)=1-p^{-2}\tilde{\Pi}_L(t,\textbf{p};\beta)=1+c_{TQED}\ \omega(t;\beta),
\end{equation}
and,
\begin{equation}\label{mag}
  \mu^{-1}(t;\beta)=1+p^{-2}\tilde{\Pi}_T(t,\textbf{p};\beta)-p^{-4}p_0^2\Pi_L(t,\textbf{p};\beta)=1-c_{TQED}\ \omega(t;\beta)\ .
\end{equation}
However, the phase velocity obviously remains smaller than the speed of light $c$; this is formally related to the fact that the boundary conditions on the fluctuating internal photons, which are considered as thermalized \cite{Weldon1982}:
\begin{equation}\label{speed}
  \upsilon_p(t;\beta)=\sqrt{\frac{\mu^{-1}(t;\beta)}{\epsilon(t;\beta)}}\simeq 1-c_{TQED}\ \omega(t;\beta).
\end{equation}
The photon-photon scattering is studied at low energy using the effective field theory where all fermionic degrees of freedom have been integrated out. In the low-energy and/or low-temperature sector, smaller than the electron mass of QED, the only relevant degrees of freedom in the effective theory will be photons. Consequently, the effective Lagrangian of the theory should be constructed from the photon field with vertices describing the non-linear self-interactions of the electromagnetic field, fundamentally could be mediated by any closed charged fermion loop. In the original theory, only electron loops are considered due to their small mass.
The resulting theory, often referred to as the Euler Heisenberg (EH) effective theory, admits a coupling of $2n$ photons with one another, where $n\geq2$. Therefore, we consider the photon polarization tensor in an effective theory for photon-photon interaction. The energies and momenta of the photons are so small $(p_0,p\ll m)$, that pair creation or annihilation are impossible. These photons can interact with each other via photon-photon scattering, which is described by the two-dimension eight operators in effective Lagrangian for the photon-photon interaction. And also, we get the components of the polarization by tracing over one pair of indices in the four vertex (please see after equation 19). Then the photon polarization tensor $\Pi_{\mu\nu}$ describes the propagation of photons in 4-dimensional space-time and carries the information of polarization.

Actually, our work at low temperature by using the effective field theory $\mathcal{L}(T\ll m)$. For the high temperature limit $\mathcal{L}(T\gg m)$, when the temperatures above m, the thermal bath cancels the vacuum polarization effects from the zero temperature EH Lagrangian \cite{Manjarres2017}. Therefore, to see the behavior of the speed of light at high temperature one need to start with a more complete Lagrangian
describing the photon-photon interaction at any energy.

From (\ref{ele}) and (\ref{mag}) when $\textbf{p}=0$ the longitudinal and transverse dielectric functions coincide in such limit i.e. $\varepsilon_{T,L}(t,\textbf{p=0};\beta)=1+c_{TQED}\ \omega(t;\beta)$ \cite{Weldon1982}.
Finally, we can write the quantities of the thermal QED in the mixed space.

It is very interesting to evaluate the effect of the radiative emission and absorption of particles in the heat bath on the electromagnetic properties of hot media. In fact, this is what we have obtained in the present work. The CMB has a thermal black body spectrum at low temperature as $2.7 K$ \cite{Fixsen2009}. From the fig.(\ref{figappd}), the numerical value extracted of $\omega(t=0;T)$  at a temperature of $2.7K$ agrees with results in \cite{Thoma2000,TARRACH1983}.  Consequently, the interaction of photons in a cold heat bath leads to slowed down low-frequency electromagnetic waves. For different temperatures, the relative change in the speed of light is of the order of $\omega(t=0;T)$, being the dominant thermal effect at low temperatures on the propagation of electromagnetic waves. Since the temperature of the universe decreases with time, the phase velocity increases continuously towards its vacuum value. Whereas from the fig.(\ref{figappd}), with decreasing of the temperature and when radiation and matter are decoupled, it increased today to $\upsilon(t=0;T=2.7)=1+\left(-4.9\times10^{-43}\right)$. One can notice the slight difference between our results and those obtained in \cite{Thoma2000}, which is may be due to the fact that the numerical values of the constants used in the calculations differ by the number of decimal places since the analytical formulas are exactly the same.

The very smallness of this change renders it, as a non-measurable effect, the fact remains that the speed of light is not really constant in our universe. Our result on increasing the speed of light is completely in qualitative agreement with the result obtained in the context of variable speed of light (VSL) theories. Indeed these theories obtained a divergent behavior for the two temperature limits, namely high temperature (HT) limit corresponding to the early hot universe and the low temperature (LT) limit corresponding to the ultra-cold universe. The speed of light diverges really in the mathematical sense:$ \lim\limits_{\substack{T \rightarrow LT \\ T \rightarrow HT}} c(T) = \infty $ \cite{Cruz2018}. This means that the speed of light is expected to increase as the universe approaches its final destination.

The fact that the speed of light is greater than the standard value violates, in a way, one of the consequences of the Lorentz transformation in special relativity (SR). One has tried to explain this as being due to the approximate Lagrangian within the framework of the effective field theory $\mathcal{L}(T\ll m)$, an explanation which turns out to be inconsistent because the approximated Lagrangian remains in its relativistic invariant form. Although in some cases of calculations with approximations, non-physical characteristics can appear, suggesting that the neglected terms may be of great importance. However, when we take a close look at the standard self-energy of the photon in momentum space
\begin{equation}\label{spe1}
\widetilde{\Pi}^{(\sigma,1/2)}_{T,L}(P;\beta)= \chi_{T,L} \int_{0}^{\infty}dq\frac{q^3}{e^{\beta q}-1}=\chi_{T,L}\left[6\frac{\zeta(4)}{\beta^4}\right],
\end{equation}
where $\zeta(4)$ is zeta function and compare it to the photon self energy in the mixed space,
\begin{equation}\label{spe2}
\widetilde{\Pi}^{(\sigma,1/2)}_{T,L}(t,p;\beta)= \chi_{T,L} \int_{0}^{\infty}\frac{q^3}{e^{\beta q}-1}\cos qt \ dq=\chi_{T,L}\frac{\psi^3\left(1+i\frac{t}{\beta}\right)+\psi^3\left(1-i\frac{t}{\beta}\right)}{2\beta^4},
\end{equation}
one can recognize at first sight, that the source of the violation in question is nothing that $\cos qt$-term. This particular term, which comes from the initial definition of the propagator in mixed space (see relation(\ref{propagator})), seems to lead to a very short time oscillation inducing an increase of the speed of light. Due to the structure of the Fourier transform, the formalism in mixed space apparently is not covariant.
Indeed, when we consider the case $tT\ll 1$, the thermal speed of light in the mixed space $\upsilon_p(t;T)$ (relation (\ref{speed}))is given by,

\begin{equation}\label{speedDL}
\upsilon_p(t;T)\simeq 1-\left\{\frac{44}{2025}\left(\frac{\pi\ \alpha}{ m^2}\right)^2T^4+\frac{88}{2025}\left(\frac{\pi^2\ \alpha}{m^2}\right)^2\frac{T^8}{21}\left[\frac{\pi^2}{10}t^4-6\frac{t^2}{T^2}\right]\right\},
\end{equation}

from which we recover the timeless case $t=0$,

\begin{equation}\label{speedDL0}
\upsilon_p(t=0;T)\simeq 1-\frac{44}{2025}\left(\frac{\pi\ \alpha}{ m^2}\right)^2T^4,
\end{equation}

which is in complete agreement with those obtained in different other works and without violating SR.

The study of the intimate relationship between QED vacuum structure and the propagation of light is not new in itself. This allowed us to understand why the propagation of photon in non-trivial QED vacuum is very different from that in normal one. A different behavior which is due to the interactions between photons and the structure of the modified vacuum. In our case, the light-medium interaction is a quantum effect, is caused by the phenomenon of vacuum polarization and makes the photons superluminal. This effect is obtained from a calculation taking into account one-loop correction to the usual QED propagator in the mixed space and using the symmetric propagator. The obtained deviation from the standard speed of light is of the order $| \upsilon_p(t;T)-1 |\sim 10^{-45}$ at LT which is in complete agreement with the order of all deviations obtained from previous works \cite{Scharnhorst1990,Barton1990,Drummond1980,Daniels1994,Shore1996}. Indeed, from these works and in order to calculate the speed of photons in a QED vacuum modified by certain external conditions, deviations from the standard speed of light were obtained. These deviations are specific to the propagation of photons in very particular environments such as the curved space-time vacuum \cite{Drummond1980}, the Casimir vacuum with boundary conditions \cite{Scharnhorst1990,Barton1990} and the thermal vacuum\cite{TARRACH1983}. The speed of the photons is greater than the standard speed of light in curved space-time and Casimir media and is always lower in other cases. All the deviations reported in the different articles, including our result, are perturbative with a very small numerical value, making their detection very difficult, if not impossible. Moreover, from the wave point of view, the detection of a wave with an $\omega \ll m$  is another practically impossible task. The fundamental question, whether one can measure such deviations in the speed of light at low energy, remains open.

In the case, where the situation is considered to be close to ours, namely Casimir vacuum, the speed of the photons is superluminal. The work was firstly performed by K. Scharnhorst \cite{Scharnhorst1990} and rederived by G. Barton \cite{Barton1990}. Then came another calculation, based on a slightly different approach, performed by S. Ben-Menahem\cite{Ben-Menahem1990}, using causal graphs rather than Feynman graphs, from which he showed that the wavefront was moving at exactly the standard speed of light, so without any problem with relativity. Finally, one can say that our superluminal photons, in addition to the deviation of their speed which is extremely small beyond any experimental attempt, one can summarize two main reasons of having them: 1) The use of the symmetric propagator instead the causal one. 2) The use of one loop approximation instead two loop or beyond.
Despite this, we can underline that this phenomenon could generate a questioning of something, or a novelty, in the quantum field theory  with the interaction due to the polarization of the vacuum.
A quantum field-theoretical explanation in terms of modes suggests the following physical picture of why photons move faster between plates than in a normal vacuum, in contrast to what happens in a thermal vacuum. Modifications of the thermal vacuum that populate it with real or virtual particles introduce coherent scattering which reduces the speed of photons. However, in the Casimir vacuum, the modifications induce the disappearance of some virtual modes and, consequently, the rate of scattering. The speed of photons is, then, increased. For this reason one can imagine the following statement : whatever the created vacuum, it is not so empty, so the standard speed of light that we measure is only that of photons in a vacuum with a certain polarization, even if it is small. Therefore, if the experimental techniques improve to have a vacuum less and less modifiable, the speed of light is expected to increase, because in a less modifiable vacuum there is less scattering.

Another attempt at a plausible explanation is to say that there is really no violation of relativistic invariance since we are dealing with photons, massless particles and not massive particles. Indeed, the only fundamental constant present in SR is the speed of light, the ultimate speed which is impossible to achieve by any massive particle. If left to vary, it will not violate the principles of SR theory. Thus, when writing the function $\upsilon_p(t;T)$, the invariance of the speed of light must only be considered with respect to the motion of particles of non-zero mass at rest, but not with respect to time and temperature. For the general motion of massive particles, the value $\upsilon_p(t;T)$ for a given temperature and time $(T,t)$ will remain the limiting speed which is invariant in the framework of SR.
Finally, the study with a complete Lagrangian, without approximation, is necessary. A study which is likely to be more complicated, will take more time and can be explored in a future work.

Also, we can note that this result agrees with the prediction of another model based on the generalized uncertainty principle. Indeed, a modified velocity of photons is obtained, indicating that photons propagation depends on their energy, providing possible values greater than the vacuum light velocity \cite{Majhi2013}. A different energy dispersion relation is predicted by some approaches to quantum gravity, from which one can deduce a modified light velocity \cite{Camelia1998}. A common picture of both thermal vacuum and quantum-gravitational medium emerges, responding differently to the propagation of particles of different energies, with new velocity dispersion law.
The expansion of our universe has been accompanied by adiabatic cooling, causing the energy density of the plasma to decrease until it becomes favourable for the combination of electrons with protons to form hydrogen atoms.  This recombination event happened when the temperature was around $3000 K$. The fact that there is no interaction between photons and neutral atoms, rendering them to travel freely through space and inducing the decoupling of matter and radiation \cite{Gawiser2000}.

Also we have determined the changes in the electric permittivity and the magnetic permeability besides the velocity of light via $\Omega(t;T)$ from (\ref{ele}), (\ref{mag}) and (\ref{speed}). These results in a reduced, dispersion-free velocity of light, which increases during the evolution of the universe as the temperature of the CMB drops.

\section{\textbf{CONCLUSION}}\label{six}

We have calculated the photon self-energy in a photon gas using the RTF of quantum field theory at positive temperature. For this purpose, we considered the effective Lagrangian for photon-photon interaction, and we calculated the photon self-energy to lowest-order perturbation thermal theory using an effective 4-photon vertex in mixed-space by exploiting the recognized simplifications provided by this space compared to the momentum space.
Based on the arbitrary parameter $\sigma$, a general path integral description is considered. Our starting point was to derive the general propagator of the scalar field in the new basis (\ref{rraa}), that allowed us to compute the photon polarization tensor in RTF. A family of paths is used in our computation in the mixed space (\ref{rraamix}), and when we choose the closed time path formalism, we obtained the same results as in the $R/A$ basis \cite{Thoma2000}. Besides the RTF approach, our work was derived in the mixed space, thus the temporal part $e^{-it q_0}$ and the thermal part $e^{\sigma \beta q_0}$ appear, both of them, in the calculations, giving rise to the function depending on the complex time $\tau$.

Indeed, photons interaction at low temperature has been taken as an application with the very known CMB radiation. Note that the photon self-energy is the same for both approaches in RTF (\ref{mixlall}), since it is related only by the symmetric propagator. However, the other components depend on the $\sigma$ parameter.
We have evaluated the changes in the electromagnetic properties of hot media caused by the radiative emission and absorption of particles in the heat bath. In fact, a fluctuation in the numerical values of the different quantities considered in this work is obtained during a small time and temperature values. Such variations are in good agreement with the predictions from other approaches \cite{Cruz2018}.

\section*{Acknowledgments}
The first author would like to thank Markus. H. Thoma and Ashok. Das for their discussion and comments. This article was supported by the Fundamental Research Grant Scheme (FRGS) under the Ministry of Higher Education with project number FRGS/1/2019/STG02/UPM/02/3.

\appendix
\section{}\label{a}
The effective photon vertex for four photons in QED was presented in Eq. (\ref{selfenergy}). Our four-point vertex is extracted from the original one given in Ref. \cite{Halter1993}; the details of the calculations are shown below.\\
$\textcolor[rgb]{0.00,0.07,1.00}{\mathcal{T}^{\alpha\beta\mu\nu}(P_1,P_2,P_3,P_4)}=g^{\alpha\beta}g^{\mu\nu}(-32i a_1 P_1.P_2P_3.P_4-8i a_2P_1.P_3P_2.P_4-8ia_2P_1.P_4P_2.P_3)\\
+g^{\alpha\mu}g^{\beta\nu}(-32ia_1P_1.P_3P_2.P_4-8ia
_2P_1.P_2P_3.P_4-8ia_2P_1.P_4P_2.P_3)\\
+g^{\alpha\nu}g^{\beta\mu}(-32ia_1P_1P_4P_2.P_3-8ia_2P_1.P_2P_3.P_4-8ia_2P_1.P_3P_2.P_4)\\-32ia_1(P_1^\mu P_2^\nu P_3^\alpha P_4^\beta+P_1^\nu P_2^\mu P_3^\beta P_4^\alpha+P_1^\beta P_2^\alpha P_3^\nu P_4^\mu)-8ia_2(P_1^\mu P_2^\nu P_3^\beta P_4^\alpha+P_1^\mu P_2^\alpha P_3^\nu P_4^\beta+P_1^\nu P_2^\alpha P_3^\beta P_4^\mu+P_1^\nu P_2^\mu P_3^\alpha P_4^\beta+P_1^\beta P_2^\nu P_3^\alpha P_4^\mu+P_1^\beta P_2^\mu P_3^\nu P_4^\alpha)\\
-g^{\alpha\beta}[-32ia_1P_3^{\nu}P_4^{\mu}P_1.P_2+8ia_2(2P_3.P_4(P_1^{\mu}P_2^{\nu}+P_1^{\nu}P_2^{\mu})-P_1^{\mu}P_3^{\nu}P_2.P_4-P_2^{\nu}P_4^{\mu}P_1.P_3-P_2^{\mu}P_3^{\nu}P_1.P_4-P_1^{\nu}P_4^{\mu}P_2.P_3)]\\
-g^{\alpha\mu}[-32ia_1P_2^{\nu}P_4^{\beta}P_1.P_3+8ia_2(2P_2.P_4(P_1^{\nu}P_3^{\beta}+P_1^{\beta}P_3^{\nu})-P_1^{\beta}P_2^{\nu}P_3.P_4-P_3^{\nu}P_4^{\beta}P_1.P_2-P_2^{\nu}P_3^{\beta}P_1.P_4-P_1^{\nu}P_4^{\beta}P_2.P_3)]\\
-g^{\alpha\nu}[-32ia_1P_2^{\mu}P_3^{\beta}P_1.P_4+8ia_2(2P_2.P_3(P_1^{\mu}P_4^{\beta}+P_1^{\beta}P_4^{\mu})-P_3^{\beta}P_4^{\mu}P_1.P_2-P_2^{\mu}P_4^{\beta}P_1.P_3-P_1^{\mu}P_3^{\beta}P_2.P_4-P_1^{\beta}P_2^{\mu}P_3.P_4)]\\
-g^{\mu\nu}[-32ia_1P_1^{\beta}P_2^{\alpha}P_3.P_4+8ia_2(2P_1.P_2(P_3^{\alpha}P_4^{\beta}+P_3^{\beta}P_4^{\alpha})-P_1^{\beta}P_4^{\alpha}P_2.P_3-P_1^{\beta}P_3^{\alpha}P_2.P_4-P_2^{\alpha}P_4^{\beta}P_1.P_3-P_2^{\alpha}P_3^{\beta}P_1.P_4)]\\
-g^{\mu\beta}[-32ia_1P_1^{\nu}P_4^{\alpha}P_2.P_3+8ia_2(2P_1.P_4(P_2^{\nu}P_3^{\alpha}+P_2^{\alpha}P_3^{\nu})-P_2^{\nu}P_4^{\alpha}P_1.P_3-P_3^{\nu}P_4^{\alpha}P_1.P_2-P_1^{\nu}P_3^{\alpha}P_2.P_4-P_1^{\nu}P_2^{\alpha}P_3.P_4)]\\
-g^{\nu\beta}[-32ia_1P_1^{\mu}P_3^{\alpha}P_2.P_4+8ia_2(2P_1.P_3(P_2^{\mu}P_4^{\alpha}+P_2^{\alpha}P_4^{\mu})-P_2^{\mu}P_3^{\alpha}P_1.P_4-P_1^{\mu}P_2^{\alpha}P_3.P_4-P_3^{\alpha}P_4^{\mu}P_1.P_2-P_1^{\mu}P_4^{\alpha}P_2.P_3)]$\\
A four-point vertex with the understanding that the net momentum flowing into the vertex is zero:\\
$\textcolor[rgb]{0.00,0.07,1.00}{\mathcal{T}^{\alpha\beta\mu\nu}(Q,P,Q,P)}=g^{\alpha\beta}g^{\mu\nu}(-32i a_1 Q.PQ.P-8i a_2Q.QP.P-8ia_2Q.PP.Q)\\
+g^{\alpha\mu}g^{\beta\nu}(-32ia_1Q.QP.P-8ia_2Q.PQ.P-8ia_2Q.PP.Q)\\
+g^{\alpha\nu}g^{\beta\mu}(-32ia_1QPP.Q-8ia_2Q.PQ.P-8ia_2Q.QP.P)\\-32ia_1(Q^\mu P^\nu Q^\alpha P^\beta+Q^\nu P^\mu Q^\beta P^\alpha+Q^\beta P^\alpha Q^\nu P^\mu)-8ia_2(Q^\mu P^\nu Q^\beta P^\alpha+Q^\mu P^\alpha Q^\nu P^\beta+Q^\nu P^\alpha Q^\beta P^\mu+Q^\nu P^\mu Q^\alpha P^\beta+Q^\beta P^\nu Q^\alpha P^\mu+Q^\beta P^\mu Q^\nu P^\alpha)\\
-g^{\alpha\beta}[-32ia_1Q^{\nu}P^{\mu}Q.P+8ia_2(2Q.P(Q^{\mu}P^{\nu}+Q^{\nu}P^{\mu})-Q^{\mu}Q^{\nu}P.P-P^{\nu}P^{\mu}Q.Q-P^{\mu}Q^{\nu}Q.P-Q^{\nu}P^{\mu}P.Q)]\\
-g^{\alpha\mu}[-32ia_1P^{\nu}P^{\beta}Q.Q+8ia_2(2P.P(Q^{\nu}Q^{\beta}+Q^{\beta}Q^{\nu})-Q^{\beta}P^{\nu}Q.P-Q^{\nu}P^{\beta}Q.P-P^{\nu}Q^{\beta}Q.P-Q^{\nu}P^{\beta}P.Q)]\\
-g^{\alpha\nu}[-32ia_1P^{\mu}Q^{\beta}Q.P+8ia_2(2P.Q(Q^{\mu}P^{\beta}+Q^{\beta}P^{\mu})-Q^{\beta}P^{\mu}Q.P-P^{\mu}P^{\beta}Q.Q-Q^{\mu}Q^{\beta}P.P-Q^{\beta}P^{\mu}Q.P)]\\
-g^{\mu\nu}[-32ia_1Q^{\beta}P^{\alpha}Q.P+8ia_2(2Q.P(Q^{\alpha}P^{\beta}+Q^{\beta}P^{\alpha})-Q^{\beta}P^{\alpha}P.Q-Q^{\beta}Q^{\alpha}P.P-P^{\alpha}P^{\beta}Q.Q-P^{\alpha}Q^{\beta}Q.P)]\\
-g^{\mu\beta}[-32ia_1Q^{\nu}P^{\alpha}P.Q+8ia_2(2Q.P(P^{\nu}Q^{\alpha}+P^{\alpha}Q^{\nu})-P^{\nu}P^{\alpha}Q.Q-Q^{\nu}P^{\alpha}Q.P-Q^{\nu}Q^{\alpha}P.P-Q^{\nu}P^{\alpha}Q.P)]\\
-g^{\nu\beta}[-32ia_1Q^{\mu}Q^{\alpha}P.P+8ia_2(2Q.Q(P^{\mu}P^{\alpha}+P^{\alpha}P^{\mu})-P^{\mu}Q^{\alpha}Q.P-Q^{\mu}P^{\alpha}Q.P-Q^{\alpha}P^{\mu}Q.P-Q^{\mu}P^{\alpha}P.Q)]$\\In the case of $Q^2=0$: \\
$\textcolor[rgb]{0.00,0.07,1.00}{\mathcal{T}^{\alpha\beta\mu\nu}(Q,P,Q,P)\Big|_{Q^2=0}}=i(g^{\alpha\beta}g^{\mu\nu}(-32 a_1 (Q.P)^2-8 a_2 (Q.P)^2)
+g^{\alpha\mu}g^{\beta\nu}(-8a_2(Q.P)^2-8a_2(Q.P)^2)
+g^{\alpha\nu}g^{\beta\mu}(-32a_1(Q.P)^2-8a_2(Q.P)^2)\\-32a_1(Q^\mu P^\nu Q^\alpha P^\beta+Q^\nu P^\mu Q^\beta P^\alpha+Q^\beta P^\alpha Q^\nu P^\mu)-8a_2(Q^\mu P^\nu Q^\beta P^\alpha+Q^\mu P^\alpha Q^\nu P^\beta+Q^\nu P^\alpha Q^\beta P^\mu+Q^\nu P^\mu Q^\alpha P^\beta+Q^\beta P^\nu Q^\alpha P^\mu+Q^\beta P^\mu Q^\nu P^\alpha)\\
-g^{\alpha\beta}[-32a_1Q^{\nu}P^{\mu}Q.P+8a_2(2Q.P(Q^{\mu}P^{\nu}+Q^{\nu}P^{\mu})-Q^{\mu}Q^{\nu}P^2-P^{\mu}Q^{\nu}Q.P-Q^{\nu}P^{\mu}P.Q)]\\
-g^{\alpha\mu}[+8a_2(2P^2(Q^{\nu}Q^{\beta}+Q^{\beta}Q^{\nu})-Q^{\beta}P^{\nu}Q.P-Q^{\nu}P^{\beta}Q.P-P^{\nu}Q^{\beta}Q.P-Q^{\nu}P^{\beta}P.Q)]\\
-g^{\alpha\nu}[-32a_1P^{\mu}Q^{\beta}Q.P+8a_2(2P.Q(Q^{\mu}P^{\beta}+Q^{\beta}P^{\mu})-Q^{\beta}P^{\mu}Q.P-Q^{\mu}Q^{\beta}P^2-Q^{\beta}P^{\mu}Q.P)]\\
-g^{\mu\nu}[-32a_1Q^{\beta}P^{\alpha}Q.P+8a_2(2Q.P(Q^{\alpha}P^{\beta}+Q^{\beta}P^{\alpha})-Q^{\beta}P^{\alpha}P.Q-Q^{\beta}Q^{\alpha}P^2-P^{\alpha}Q^{\beta}Q.P)]\\
-g^{\mu\beta}[-32a_1Q^{\nu}P^{\alpha}P.Q+8a_2(2Q.P(P^{\nu}Q^{\alpha}+P^{\alpha}Q^{\nu})-Q^{\nu}P^{\alpha}Q.P-Q^{\nu}Q^{\alpha}P^2-Q^{\nu}P^{\alpha}Q.P)]\\
-g^{\nu\beta}[-32a_1Q^{\mu}Q^{\alpha}P^2+8a_2(P^{\mu}Q^{\alpha}Q.P-Q^{\mu}P^{\alpha}Q.P-Q^{\alpha}P^{\mu}Q.P-Q^{\mu}P^{\alpha}P.Q)])\\$
\unskip
\section{}\label{b}
\subsection{}\label{b0}
The eqs.(\ref{rraa}) and (\ref{rraamix}) when $\varphi=\frac{\pi}{4}$ the basis will be write as:
\begin{equation*}
  \begin{array}{l}
\tilde{D}^{(\sigma,11)}(Q;\beta)=\frac{\pi\delta(Q^2)}{i}\left[1-\left(\Theta(-q_0)e^{\sigma\beta q_0}+\Theta(q_0)e^{-\sigma\beta q_0}\right)+2n_B(Q;\beta)\left[1-\cosh\sigma\beta q_0\right]\right],\\
\tilde{D}^{(\sigma,21)}(Q;\beta)=\left[\frac{1}{2}\mathcal{P}\frac{1}{Q^2}+\frac{\pi\delta(Q^2)}{i}\left(\Theta(-q_0)e^{\sigma\beta q_0}-\Theta(q_0)e^{-\sigma\beta q_0}\right)+2n_B(Q;\beta)\sinh\sigma\beta q_0\right],\\
\tilde{D}^{(\sigma,12)}(Q;\beta)=\left[\frac{1}{2}\mathcal{P}\frac{1}{Q^2}+\frac{\pi\delta(Q^2)}{i}\left(\Theta(q_0)e^{-\sigma\beta q_0}-\Theta(-q_0)e^{\sigma\beta q_0}\right)-2n_B(Q;\beta)\sinh\sigma\beta q_0\right],\\
\tilde{D}^{(\sigma,22)}(Q;\beta)=\frac{\pi\delta(Q^2)}{i}\left[1+\left(\Theta(-q_0)e^{\sigma\beta q_0}+\Theta(q_0)e^{-\sigma\beta q_0}\right)+2n_B(Q;\beta)\left[1+\cosh\sigma\beta q_0\right]\right],
\end{array}
\end{equation*}
and
 \begin{equation*}
  \begin{array}{l}
\tilde{D}^{(\sigma,11)}(t,q;\beta)=\frac{\cos qt}{2iq}\left[\left(1-e^{-\sigma\beta q}\right)+2n_B(q;\beta)\left[1-\cosh\sigma\beta q\right]\right],\\
\tilde{D}^{(\sigma,21)}(t,q;\beta)=\frac{\sin qt}{2q}\left[\left(\Theta(-t)-\Theta(t)\right)+e^{-\sigma\beta q}-2n_B(q;\beta)\sinh\sigma\beta q\right],\\
\tilde{D}^{(\sigma,12)}(t,q;\beta)=\frac{\sin qt}{2q}\left[\left(\Theta(-t)-\Theta(t)\right)-e^{-\sigma\beta q}+2n_B(q;\beta)\sinh\sigma\beta q\right],\\
\tilde{D}^{(\sigma,22)}(t,q;\beta)=\frac{\cos qt}{2iq}\left[\left(1+e^{-\sigma\beta q}\right)+2n_B(q;\beta)\left[1+\cosh\sigma\beta q\right]\right].
\end{array}
\end{equation*}
\subsection{}\label{b1}
The parameter $\chi_{T,L}$ is refer to the transverse projector is $\chi_{T}$ and longitudinal projector $\chi_{L}$ as:
\begin{equation*}
  \chi_{T,L}=\begin{cases}
                                        \frac{44}{135}\left(i\frac{\alpha p}{\pi m^2}\right)^2, &  \mbox{if } \ \ \ \ \ \ \ \ the\ longitudinal\ projector, \\
                                        \\
                                        \frac{44}{135}\left(i\frac{\alpha p}{\pi m^2}\right)^2\frac{\left(p_0^2+ p^2\right)}{p^2}, &  \mbox{when }\ \ \ \  the\ transverse\ projector.\\
                                      \end{cases}
\end{equation*}
The integrations of the thermal terms without distribution function, in (\ref{seinmix}),
\begin{equation*}
 \int_{0}^{\infty}q^3\ f(q,t) \ e^{-\sigma\beta q}dq=\begin{cases}
                                         \left[\frac{6(t^4+(\sigma\beta)^4-6t^2\ \beta^2\ \sigma^2)}{((\beta\sigma)^2+t^2)^4}\right] ,&  \mbox{if} \ \ \ \ \ \ \ \ \ \ \ f(q,t)=\cos qt,\\
                                        \\
                                      -24i\ t\frac{ \sigma\beta\left((\sigma\beta)^2+t^2\right)}{\left((\sigma\beta)^2+t^2\right)^{4}} &  \mbox{when}\ \ \ \ \ \ \ f(q,t)=\sin qt.\\
                                      \end{cases}
\end{equation*}
The integrations of the thermal terms within distribution function, in (\ref{seinmix}), where $\lambda=\frac{t}{\beta}$:
$$\int_{0}^{\infty}\frac{x^{n-1} \cos^s\lambda x}{\exp^x-1}dx=\int_{0}^{\infty}\frac{x^{n-1} e^{-x}\frac{1}{2^s}(e^{i\lambda x}+e^{-i\lambda x})^s}{1-e^{-x}}dx$$\\
$$=\int_{0}^{\infty}\sum\limits_{r=0}^{s}\frac{\binom{s}{r}}{2^s}\frac{x^{n-1} e^{-x}e^{i\lambda (s-2r)x}}{1-e^{-x}}dx\\ =\sum\limits_{r=0}^{s}\frac{\binom{s}{r}}{2^s}\int_{0}^{\infty}\frac{x^{n-1} e^{-x(1-i\lambda (s-2r))}}{1-e^{-x}}dx$$\\
The form in this integration represents the poly gamma function $\psi(x)$ of order $(n-1)$ which is a meromorphic function on the complex numbers $\mathbb{C}$ and the $n$th derivative of the logarithm of the gamma function $\Gamma(n)=(n-1)!$:\\
$$\psi^{n-1}(\tau)=(-1)^{n}\int_{0}^{\infty}\frac{x^{n-1} e^{-x\tau}}{1-e^{-x}}dx=(-1)^n\Gamma(n)\sum_{k=0}^{\infty}\frac{1}{(\tau+k)^n}$$
where $\tau=\sqrt{1+\left(\frac{t}{\beta}\right)^2}\exp^{i\tan^{-1}\left(\frac{t}{\beta}\right)}$ is a complex variable, with $n>0$ Re $\tau>0$. Thus,\\
$$\int_{0}^{\infty}\frac{x^{n-1} \cos^s\lambda x}{\exp^x-1}dx=\frac{2^{-s}}{(-)^{n}}\sum\limits_{r=0}^{s}\binom{s}{r}\psi^{n-1}(1-i\lambda(s-2r))$$\\
$$\psi^{n-1}(1-i\lambda(s-2r))=\frac{(n-1)!}{(-)^{-(n)}}\sum\limits_{k=0}^{\infty}\frac{1}{(k+(1-i\lambda(s-2r)))^{n}}
=\frac{(n-1)!}{(-)^{-(n)}}\sum\limits_{k=0}^{\infty}\left(\frac{(k+1)+i\lambda(s-2r)}{(k+1)^2+\lambda^2(s-2r)^2}\right)^{n}$$
$$=\frac{(n-1)!}{(-)^{-(n+r)}}\sum\limits_{k=0}^{\infty}\frac{\sum\limits_{j=0}^{n}\binom{n}{j}(1+k)^{n}\left(\frac{i\lambda (s-2r)}{1+k}\right)^j}{((k+1)^2+\lambda^2(s-2r)^2)^{n}}=\frac{(n-1)!}{(-)^{-(n+r)}} \sum\limits_{k=0}^{\infty}\frac{\sum\limits_{j=0}^{n}\binom{n}{j}(1+k)^{n}\left(\frac{i\lambda (s-2r)}{1+k}\right)^j}{((k+1)^2+\lambda^2(s-2r)^2)^{n}}$$\\
The summation results will be real values when $j=0,2,...,2n$
$$
\int_{0}^{\infty}\frac{x^{n-1} \cos^s\lambda x}{\exp^x-1}dx=\frac{1}{2^s}\sum\limits_{r=0}^{s}\binom{s}{r}
(-)^{n+r}(n-1)! \sum\limits_{k=0}^{\infty}\frac{\sum\limits_{j=0}^{n}\binom{n}{j}(1+k)^{n}\left(\frac{i\lambda (s-2r)}{1+k}\right)^j}{((k+1)^2+\lambda^2(s-2r)^2)^{n}}
$$
And
$$
  \int_{0}^{\infty}\frac{x^{n-1} \sin^s\lambda x}{\exp^x-1}dx=\frac{1}{(2i)^s}\sum\limits_{r=0}^{s}\binom{s}{r}
(-)^{n+r}(n-1)! \sum\limits_{k=0}^{\infty}\frac{\sum\limits_{j=0}^{n}\binom{n}{j}(1+k)^{n}\left(-\frac{i\lambda (s-2r)}{1+k}\right)^j}{((k+1)^2+\lambda^2(s-2r)^2)^{n}}
$$
The poly gamma function has the series representation shown in the above formula. Also we have the Zeta function when $\cos^s(\lambda x)\longrightarrow 1$.
In our work we have had $s=1$ and $n=4$:
$$\int_{0}^{\infty}dx\frac{x^3}{e^{x}-1}\cos\left(\frac{xt}{\beta}\right)=\frac{1}{2}\frac{\psi^3\left(1+i\frac{t}{\beta}\right)+\psi^3\left(1-i\frac{t}{\beta}\right)}{\beta^4}$$
then,
\begin{equation*}
 \int_{0}^{\infty}q^3\ f(q,t) \ \frac{\cosh \sigma\beta q}{e^{\beta q}-1} dq=\begin{cases}
                                         \frac{1}{4}\frac{\left(\psi^3(\bar{\tau}+\sigma)+\psi^3(\tau+\sigma)+\psi^3(\bar{\tau}-\sigma)+\psi^3(\tau-\sigma)\right)}{\beta^4} ,&  \mbox{if} \ \ \ \ \ \ \ \ \ \ \ f(q,t)=\cos qt,\\
                                        \\
                                      \frac{1}{4i}\frac{\left(\psi^3(\bar{\tau}-\sigma)-\psi^3(\tau+\sigma)+\psi^3(\bar{\tau}+\sigma)-\psi^3(\tau-\sigma)\right)}{\beta^4} &  \mbox{when}\ \ \ \ \ \ \ f(q,t)=\sin qt.\\
                                      \\
                                      \frac{1}{2}\frac{\left(\psi^3(\bar{\tau})+\psi^3(\tau)\right)}{\beta^4} &  \mbox{when}\ \ \ \ \ \ \ f(q,t)=\cos qt \ \ \sigma=0.\\
                                      \\
                                       \frac{1}{2i}\frac{\left(\psi^3(\bar{\tau})-\psi^3(\tau)\right)}{\beta^4} &  \mbox{when}\ \ \ \ \ \ \ f(q,t)=\sin qt \ \ \sigma=0.\\
                                      \end{cases}
\end{equation*}
and
\begin{equation*}
 \int_{0}^{\infty}q^3\ f(q,t) \ \frac{\sinh \sigma\beta q}{e^{\beta q}-1} dq=\begin{cases}
                                         \frac{1}{4}\frac{\left(\psi^3(\bar{\tau}-\sigma)-\psi^3(\bar{\tau}+\sigma)+\psi^3(\tau-\sigma)-\psi^3(\tau+\sigma)\right)}{\beta^4} ,&  \mbox{if} \ \ \ \ \ \ \ \ \ \ \ f(q,t)=\cos qt,\\
                                        \\
                                      \frac{1}{4i}\frac{\left(\psi^3(\bar{\tau}-\sigma)-\psi^3(\tau-\sigma)-\psi^3(\bar{\tau}+\sigma)+\psi^3(\tau+\sigma)\right)}{\beta^4} &  \mbox{when}\ \ \ \ \ \ \ f(q,t)=\sin qt.\\
                                      \\
                                      0 &  \mbox{when}\ \ \ \ \ \ \ f(q,t)=\cos qt \ \ \sigma=0.\\
                                      \\
                                       0 &  \mbox{when}\ \ \ \ \ \ \ f(q,t)=\sin qt \ \ \sigma=0.\\
                                      \end{cases}
\end{equation*}
The components photon self-energy in (\ref{seinmix}) will be written as:\\
$\begin{array}{l}
\tilde{\Pi}_{T,L}^{(\sigma,11)}(t,p;\beta)=\frac{\chi_{T,L}}{2\beta^4} \left\{\beta^4\left[\frac{-6(t^4+(\sigma\beta)^4-6t^2\ \beta^2\ \sigma^2)}{((\beta\sigma)^2+t^2)^4}\right]+\left[\psi^3(\bar{\tau})+\psi^3(\tau)\right]\right.\\
\ \ \ \ \ \ \ \ \ \ \ \ \ \ \ \  \ \ \ \ \ \left.-\frac{1}{2}\left[\psi^3(\bar{\tau}-\sigma)+\psi^3(\bar{\tau}+\sigma)+\psi^3(\tau-\sigma)+\psi^3(\tau+\sigma)\right]\right\},\\
\\
\tilde{\Pi}^{(\sigma,21)}_{T,L}(t,p;\beta)= \frac{\chi_{T,L}}{2\beta^4}\left[-24i\beta^4\frac{ \sigma\beta\left((\sigma\beta)^2+t^2\right)t}{\left((\sigma\beta)^2+t^2\right)^{4}}
-\frac{1}{2}\left(\psi^3(\bar{\tau}-\sigma)-\psi^3(\bar{\tau}+\sigma)-\psi^3(\tau-\sigma)+\psi^3(\tau+\sigma)\right)\right],\\
\\
\tilde{\Pi}^{(\sigma,12)}_{T,L}(t,p;\beta)=\frac{\chi_{T,L}}{2\beta^4}\left[24i\beta^4\frac{ \sigma\beta\left((\sigma\beta)^2+t^2\right)t}{\left((\sigma\beta)^2+t^2\right)^{4}}+\frac{1}{2}\left(\psi^3(\bar{\tau}-\sigma)-\psi^3(\bar{\tau}+\sigma)-\psi^3(\tau-\sigma)+\psi^3(\tau+\sigma)\right)\right],\\
\\
\tilde{\Pi}^{(\sigma,22)}_{T,L}(t,p;\beta)=\frac{\chi_{T,L}}{2\beta^4} \left\{\left[\beta^4\frac{6(t^4+(\sigma\beta)^4-6t^2\ \beta^2\ \sigma^2)}{((\beta\sigma)^2+t^2)^4}\right]+\left[\psi^3(\bar{\tau})+\psi^3(\tau)\right]\right.\\
\ \ \ \ \ \ \ \ \ \ \ \ \ \ \ \  \ \ \ \ \ \left.+\frac{1}{2}\left[\psi^3(\bar{\tau}-\sigma)+\psi^3(\bar{\tau}+\sigma)+\psi^3(\tau-\sigma)+\psi^3(\tau+\sigma)\right]\right\}.\\
\end{array}$\\
On the way to calculating the integration of symmetric propagator in (\ref{Sp}) this case,
$$\int_{0}^{\infty}dx\frac{x^3}{e^{x}-1}\cos\left(\frac{xt}{\beta}\right)=\sum_{k=0}^{\infty}\frac{\left(1+4 k+6 k^2+k^4+4 k^3-\frac{6 k^2 t^2}{\beta ^2}-\frac{12 k t^2}{\beta ^2}+\frac{t^4}{\beta ^4}-\frac{6 t^2}{\beta ^2}\right)}{\left(1+2 k+k^2+\frac{t^2}{\beta ^2}\right)^4},$$
$$\approx-\frac{\beta^4}{t^4}+\frac{2}{3}\pi^4coth^2\left(\pi\frac{t}{\beta}\right)\ csch^2\left(\pi\frac{t}{\beta}\right)
+\frac{2}{3}\pi^4\ 2csch^4\left(\pi\frac{t}{\beta}\right).$$
this result used to construct $\tilde{\Pi}^{(\sigma,S)}_{T,L}(t,p;\beta)$.


\end{document}